\documentclass[twocolumn,pre,floatfix]{revtex4-1}
\usepackage{fancyhdr,latexsym}
\usepackage{times,graphicx,subfigure}
\usepackage{amsmath,amssymb,amsfonts}
\usepackage[sort&compress]{natbib}
\def\img{i}
\def\ee{e}

\newcommand{\ave}[1]{\ensuremath \left\langle #1 \right\rangle}

\def\ss{\underline{s}}

\def\rv{{\bf r}}
\def\kv{{\bf k}}
\def\qv{{\bf q}}

\def\Rv{{\bf R}}

\def\barchia{\bar{\chi}_{a}}

\def\Sc{S}

\def\Ss{\Omega}
\def\Ssi{\tilde{\Ss}}
\def\Sssum{\Ss}

\def\ss{\omega}
\def\ssi{\tilde{\ss}}

\def\Ch{C}

\def\Gr{G}

\def\Gh{\Gr}

\begin{document}
% -----------------------------------------------

% title page
\title{Renormalized One-loop Theory of Correlations in Disordered Diblock Copolymers}

\author{Jian Qin, Piotr Grzywacz, and David C. Morse}

\affiliation{
Department of Chemical Engineering and Materials Science, 
University of Minnesota, 
421 Washington Ave. S.E., Minneapolis, MN 55455}

\date{\today}

\begin{abstract}
A renormalized one-loop theory (ROL) \cite{Piotr_Morse_2007} is used 
to calculate corrections to the random phase approximation (RPA) for 
the structure factor $\Sc(q)$ in disordered diblock copolymer melts.
Predictions are given for the peak intensity $S(q^{\star})$, peak position
$q^{\star}$, and single-chain statistics for symmetric and asymmetric
copolymers as functions of $\chi N$, where $\chi$ is the Flory-Huggins
interaction parameter and $N$ is the degree of polymerization. The ROL and
Fredrickson-Helfand (FH) theories are found to yield asymptotically
equivalent results for the dependence of the peak intensity $S(q^{\star})$ 
upon $\chi N$ for symmetric diblock copolymers in the limit of strong 
scattering, or large $\chi N$, but yield qualitatively different 
predictions for symmetric copolymers far from the ODT and for asymmetric 
copolymers.  The ROL theory predicts a suppression of $S(q^\star)$ and a 
decrease of $q^{\star}$ for large values of $\chi N$, relative to the RPA 
predictions, but an enhancement of $S(q^{\star})$ and an increase in 
$q^{\star}$ for small $\chi N$ ($\chi N < 5$). By separating intra- and 
inter-molecular contributions to $S^{-1}(q)$, we show that the 
decrease in $q^{\star}$ near the ODT is caused by the $q$-dependence 
of the intermolecular direct correlation function, and is unrelated 
to any change in single-chain statistics, but that the increase in 
$q^{\star}$ at small values of $\chi N$ is a result of non-Gaussian 
single-chain statistics. 
\end{abstract}

\maketitle

%-------------------------------------------------------------------------------
\section{Introduction}
\label{sec:Introduction}
%-------------------------------------------------------------------------------

Disordered diblock copolymer melts exhibit composition fluctuations that 
can be measured by small angle X-ray (SAXS) and neutron (SANS) scattering 
experiments. Results of such experiments are often analyzed by fitting 
the scattering intensity as a function of wavenumber $q$ to Leibler's 
random-phase approximation (RPA) theory for the structure factor $S(q)$.
\cite{Leibler_1980} The RPA provides a rather accurate description of 
$S(q)$ in melts of very long copolymers far from the order-disorder 
transition (ODT), but fails near the ODT of nearly symmetric diblock 
copolymer, where strong composition fluctuations cause a breakdown of 
the underlying self-consistent field (SCF) approximation.\cite{Fredrickson_Helfand_1987}

The scattering intensity measured by small angle scattering from AB diblock
copolymer melts is proportional to the structure function $S(q) = \int d\rv
\langle \delta c_{A}(\rv) \delta c_{A}(0) \rangle e^{i\qv\cdot\rv}$, where
$\delta c_{A}(\rv)$ represents a deviation of the number concentration of A
monomers from its spatial average, and $q \equiv |\qv|$. Scattering from a 
disordered diblock copolymer melt generally exhibits a maximum $S(q^{\star})$ 
at a nonzero wavenumber $q^{\star}$.

Leibler's RPA theory predicts an inverse structure function of the form
\begin{equation}
   c N S_{0}^{-1}(q) = F_{0}(q R) - 2 \chi N.
   \label{SInvRPA}
\end{equation}
Here, $\chi$ is an effective interaction parameter, $N$ is the 
degree of polymerization, and $c$ is the monomer number concentration. 
The dimensionless function $F_{0}(q R)$ has a minimum at a wavenumber 
$q_0 = x_{0}/R$, yielding a corresponding maximum in $S_{0}(q)$, where 
$R \propto \sqrt{N}$ is the copolymer radius of gyration and $x_{0}$ 
is a dimensionless number that depends upon the copolymer composition, 
but that does not depend on $\chi$. Here and hereafter, we use
$S_{0}(q)$ to denote the RPA approximation for $S(q)$, and $q_{0}$ to 
denote the RPA prediction for $q^{\star}$.

Eq. (\ref{SInvRPA}) has been found to adequately describe 
the $q$-dependence of most SANS and SAXS experiments,
\cite{Bates_1985, Owens_Russel_1989} and has been widely used to 
extract values for the interaction parameter $\chi(T)$ as a function 
of temperature $T$.
\cite{Bates_Hartney_1985, Owens_Russel_1989, Mori_Hashimoto_1996, 
Maurer_Bates_Lodge_1998}
Careful comparisons of eq. (\ref{SInvRPA}) to both 
experimental and simulation results have, however, also revealed 
some limitations, particularly near the ODT: 

The RPA predicts an inverse peak intensity 
\begin{equation}
   c N S_{0}^{-1}(q_{0}) = 2[ (\chi N)_{s} - \chi N ]
   \label{SinvRPAPeak}
\end{equation}
that depends linearly on $\chi$. Here
$(\chi N)_{s} = F_{0}(q_{0}R)/2$ is the mean-field spinodal value.
For symmetric diblock copolymers, $(\chi N)_{s} = 10.495$. To make
a meaningful comparison of the RPA to a SANS or SAXS experiment,
$\chi$ must be allowed to be a function of temperature $T$, which 
is usually fit to the form $\chi(T) \simeq A/T + B$,
but is assumed to be independent of chain length $N$ and wavenumber
$q$.
%(If $\chi$ were allowed to be an arbitrary function of both 
%$T$ and $N$, the theory would cease to have either predictive 
%power or intellectual coherence).
Sufficiently far away from the ODT, the dependence of SANS and SAXS data
on $T$ and $N$ is often adequately described by the resulting theory.
In data taken near the ODT, however, plots of the inverse peak
scattering intensity vs. $1/T$ exhibit a characteristic nonlinearity
that cannot be described by the RPA.
\cite{Bates_Rosedale_1988,Bates_Rosedale_1990,Rosedale_Bates_1995,Almdal_Bates_1996,Sakamoto_Hashimoto_1995,Hashimoto_Han_1994,Bartels_Stamm_1994,Owens_Russel_1989,Wolff_Ruland_1993}
%symmetric PEP-PEE (poly(ethylene-propylene)-poly(ethylethylene))
%\cite{Bates_Rosedale_1988,Bates_Rosedale_1990}; 
%asymmetric PEP-PEE\cite{Almdal_Bates_1992}; 
%symmetric PE-PEE (poly(ethylene)-poly(ethylethylene)) 
%and PE-PEP
%(poly(ethylene)-poly(ethylene-propylene))
%\cite{Rosedale_Bates_1995}; 
%nearly symmetric low molecular weight PEP-PDMS
%(poly(ethylene-propylene)-poly(dimethylsiloxane))
%\cite{Almdal_Bates_1996}; 
%nearly symmetric low molecular weight PS-PI (poly(styrene)-poly(isoprene))
%\cite{Sakamoto_Hashimoto_1995, Hashimoto_Han_1994}; 
%symmetric PS-PpMS
%(poly(styrene)-poly(paramethylstyrene))
%\cite{Bartels_Stamm_1994}; 
%symmetric low molecular weight PS-PI and
%PS-PB(poly(styrene)-poly(1,2-butadiene))
%\cite{Owens_Russel_1989}; 
%symmetric PS-PB
%\cite{Wolff_Ruland_1993}. 
%The molecular weights covered in these studies range from a few 
%kg/mol \cite{Almdal_Bates_1996} to hundreds.\cite{Bartels_Stamm_1994}

The RPA predicts a peak wavenumber $q^{\star} = q_{0}$ that is 
proportional to the inverse radius of gyration of a Gaussian chain. 
The peak wavenumber $q^{\star}$ is thus expected to vary with chain length 
$N$ as $q^{\star} \propto N^{-1/2}$ at fixed temperature. Studies of 
homologous series of diblocks of differing $N$ at fixed $T$ have
found $q \propto N^{-1/2}$  far from the ODT, but a significantly 
stronger $N$-dependence near the ODT.
\cite{Almdal_Bates_1990, Papadakis_Posselt_1996}
In any single sample, the RPA suggests that $q_{0}$ should depend on
temperature only as a result of the slight intrinsic temperature 
dependence of the monomer statistical segment lengths. Numerous 
experiments have shown that $q^{\star}$ decreases with decreasing 
temperature at a rate that systematically increases near the ODT, 
\cite{Bates_Hartney_1985, Bates_Rosedale_1990, Almdal_Bates_1992, 
Rosedale_Bates_1995, Almdal_Bates_1996, Maurer_Bates_Lodge_1998, 
Owens_Russel_1989, Stuhn_Albrecht_1992, Wolff_Ruland_1993, 
Bartels_Stamm_1994, Sakamoto_Hashimoto_1995, Mori_Hashimoto_2001} 
and that appears to be too large to be explained by the observed 
temperature dependence of the pure component statistical segments 
lengths.\cite{Rosedale_Bates_1995, Almdal_Bates_1996}  

Simulation studies have provided analogous results for both the 
peak intensity $S(q^{\star})$ and the peak wavenumber $q^{\star}$, 
though for chains shorter than those studied in most experiments.  
Both lattice Monte Carlo and molecular dynamics simulations have 
shown a strongly non-linear dependence of $S^{-1}(q^\star)$ on 
inverse temperature,
\cite{Binder_Fried_1991a,Hoffmann_Blumen_1997a, Grest_Kremer_1999,
Matsen_Vassiliev_2003} a gradual decrease in $q^\star$ with 
decreasing temperature, \cite{Binder_Fried_1991a, Molina_Freire_1994,
Hoffmann_Blumen_1997a, Grest_Kremer_1996, Grest_Kremer_1999,
Matsen_Vassiliev_2003} and non-Gaussian chain statistics.
\cite{Binder_Fried_1991b} 
Though the shift in $q^{\star}$ near the ODT was initially described 
as a result of ``chain stretching",\cite{Almdal_Bates_1990} Binder 
and Fried \cite{Binder_Fried_1991b} found that the temperature 
dependence of $q^{\star}$ in lattice Monte Carlo simulations was 
significantly stronger than the temperature dependence of the 
radius of gyration, and so suggested that the shift in $q^{\star}$ 
might be caused primarily by changes in intermolecular correlations. 
Similar results were later obtained in an analogous experimental 
comparison by Bartels and Mortensen, \cite{Bartels_Mortensen_1995} 
in which $S(q)$ and $R_{g}$ were measured independently.

Leibler originally presented the RPA for $S(q)$ in diblock copolymer 
melts as part of the more general analysis of the weak-segregation 
limit of the self-consistent field theory (SCFT) for such systems.
\cite{Leibler_1980} 
Leibler found that SCFT predicts a second-order (continuous) transition 
between the disordered and lamellar phase in the special case of 
symmetric diblock copolymers, and a weakly first order transition for 
nearly symmetric copolymers. 
%
%In his concluding remarks,  Leibler \cite{Leibler_1980} made some 
%prescient comments about the limitations of SCFT near the critical
%point that he had predicted for symmetric diblock copolymers. 
Leibler also noted, however, that Brazovski\u{\i}\cite{Brazovskii_1975} 
had previously analyzed a phenomenological model of transitions 
between a homogeneous liquid and a periodic state in systems for 
which Landau theory predicts a second-order transition, and had 
concluded that such systems actually always exhibit a 
fluctuation-induced first-order transition. Brazovski\u{\i} also 
presented a theory for the dominant corrections to the Landau 
(or RPA) theory for $S(q)$ near such transitions, which were found 
to be unusually large for this class of system.  It was thus clear 
from the outset of interest in this subject that a more sophisticated 
treatment of fluctuation effects would be needed to adequately 
describe the vicinity of the ODT.

\subsection{Fredrickson-Helfand Theory}

The task of incorporating fluctuation effects into Leibler's theory was
undertaken by Fredrickson and Helfand.\cite{Fredrickson_Helfand_1987}
Fredrickson and Helfand approached the problem by constructing an
approximate mapping of Leibler's theory for block copolymer melts onto 
Brazovski\u{\i}'s theory of weakly first order crystallization.
\cite{Brazovskii_1975} 
The Brazovski\u{\i} and FH theories are both based upon an expression 
for the partition function $Z$ as a functional integral of a composition 
order parameter field $\psi(\rv)$, of the form
\begin{equation}
   Z = \int D[\psi] \; e^{-H[\psi]/k_BT}.
   \label{Zpsi}
\end{equation}
Fredrickson and Helfand began their analysis by assuming that the
effective Hamiltonian in eq. (\ref{Zpsi}) can be approximated by the 
SCFT free energy functional, while using Leibler's Taylor expansion
of this quantity. We will refer to this assumption as a mean-field 
effective Hamiltonian (MFEH) approximation. They then introduced a 
variety of further mathematical approximations in order to map the 
SCFT free energy functional to the relatively simple 
phenomenological expression considered by Brazovski\u\i.

The FH theory yields a prediction for the inverse structure factor
$S^{-1}(q)$ as a sum
\begin{equation}
   S^{-1}(q) = S_{0}^{-1}(q) + \delta S^{-1}(q),
\end{equation}
where $S_{0}(q)$ is the RPA structure factor, eq. (\ref{SInvRPA}), 
and where $\delta S^{-1}(q)$ is given by a self-consistent equation
\begin{equation}
    cN \delta S^{-1}(q) = 
    \frac{1}{\bar{N}^{1/2}}
    \frac{B}{\sqrt{\tau}},
    \label{dSinvFH}
\end{equation}
in which
\begin{equation}
   \tau \equiv c N S^{-1}(q^{\star})
   \label{dSinvPeakRPA}
\end{equation}
is a normalized inverse peak intensity, and
\begin{equation}
   \bar{N} \equiv N (cb^{3})^{2}  
   \label{NbarDef}
\end{equation}
is the so-called invariant degree of polymerization. 

The parameter $\bar{N}^{1/2}$ that appears in eq. (\ref{dSinvFH}) is a 
measure of chain overlap: $\bar{N}^{1/2}$ is proportional to the number 
of chains, each occupying a volume $N/c$, that can fit in the volume 
$R^{3} \sim N^{3/2}b^{3}$ explored by any one chain.  Values of $\bar{N}$ 
in experimental studies on diblock copolymer melts near the ODT have 
ranged from about 500 (ref. \cite{Sakamoto_Hashimoto_1995}) to 
$10000$ (ref. \cite{Rosedale_Bates_1995}), corresponding to systems 
with molecular weights that range approximately from 10 to 200 kg/mol. 

The mathematical approximations used by Fredrickson and Helfand 
yield an expression for $\delta S^{-1}(q)$ that is independent of 
$q$, as indicated in eq. (\ref{dSinvFH}). This yields a peak wavenumber
$q^{\star}$ equal to that predicted by the RPA. A subsequent extension of 
the theory by Barrat and Fredrickson \cite{Barrat_Fredrickson_1991} 
(BF) allowed for the possibility of a fluctuation-induced shift in 
$q^{\star}$, and predicted a peak wavenumber $q^{\star}$ that decreases with 
increasing $\chi N$ near the ODT, as seen in experiments and simulations.

Several experimental studies have quantitatively compared data for
both the peak intensity and $S(q^{\star})$ and the peak wavenumber $q^{\star}$
for model diblock copolymer melts near the ODT to the FH and 
BF theories. The FH theory has been found to 
describe the temperature dependence of the peak intensity near the 
ODT reasonably well in several systems.
\cite{Bates_Rosedale_1988,Wolff_Ruland_1993} Almdal {\it et al.} 
\cite{Almdal_Bates_1990} have also compared the $N$ dependence of 
$q^{\star}$ in a series of symmetric diblock copolymers at constant 
temperature to the predictions of the BF theory, 
and also reported reasonable quantitative agreement.  
%These comparisons all assumed the existence of an $N$-independent 
%interaction parameter with a temperature dependence of the form 
%$\chi(T) = A/T + B, in which $A$ and $B$ are treated as fitting 
%parameters. 

Notwithstanding its success in describing many aspects of the
experimental results, the FH theory has several shortcomings that are
inherent in how it was derived\cite{Morse_Qin_2011}:

(1) The use of the SCF free energy functional as an effective
Hamiltonian (the MFEH approximation) has no rigorous basis.  

(2) Predictions of both the full one-loop MFEH theory and the 
approximation studied by Fredrickson and Helfand are very sensitive 
to the effects of short-wavelength (monomer scale) fluctuations 
that these coarse-grained theories cannot accurately describe. 
In the jargon of field theory, the theory is ultraviolet (UV) 
divergent.

(3) Mathematical simplifications introduced by Fredrickson and 
Helfand limit the potential range of validity of the theory to 
wavenumbers $q \sim q^{\star}$ at temperatures very near the ODT 
in melts of long, nearly symmetric diblock copolymers. 

The UV divergence of the FH theory is a complication that was not 
necessarily fatal, but that was not (we think) taken sufficiently 
seriously in early work on this subject. Fredrickson and Helfand 
simply ignored all UV divergent contributions to their expression 
to $\delta S^{-1}(q)$ without commenting on the existence or 
possible physical interpretation of the divergence, and simply
reported the UV convergent parts of the resulting integrals.
Such a procedure can generally be justified if and only if 
it can be shown that the neglected terms can be absorbed 
into a renormalization of the values of few phenomenological 
parameters, such as the $\chi$ parameter. Kudlay and
Stepanow \cite{Kudlay_Stepanow_2003} later analyzed the 
UV divergent contributions in the MFEH theory for $\delta S^{-1}(q)$, 
and asked whether they could be absorbed into a simple 
renormalization of the value of the interaction parameter $\chi$, 
but concluded that this interpretation was untenable. 

\subsection{Renormalized Auxiliary Field Theory}

More recently, several authors have contributed to the development 
of a set of very closely related renormalized auxiliary field theory of corrections to the RPA 
\cite{Wang_2002,Beckrich_Wittmer_2007,Wittmer_Beckrich_2007,
Piotr_Morse_2007,Qin_Morse_2009,Morse_Qin_2011} that has both a 
more rigorous theoretical basis and a potentially wider range of 
validity than the FH and other MFEH theories. We will refer to 
this as a renormalized one-loop (ROL) theory.  The present paper 
presents predictions of the ROL theory developed in ref. 
\cite{Piotr_Morse_2007} for correlations in disordered diblock 
copolymer melts.

The way that the ROL theory is derived avoids most of the aforementioned
limitations  of the FH theory:

(1) The theory has a rigorous starting point: It is based on the
Edwards auxiliary field representation of the partition function
$Z$ for a simple coarse-grained model. 

(2) It has been shown\cite{Piotr_Morse_2007} that the UV divergence of the one-loop 
auxiliary field theory for $S(q)$ can be removed by a renormalization
procedure in which all contributions that are sensitive to monomer 
scale structures are absorbed into shifts in the values of a few
phenomenological parameters, {\it i.e.}, of the effective interaction 
parameter $\chi$ and the monomer statistical segment lengths.

(3) The range of validity of the theory is not intrinsically limited 
to the vicinity of the ODT, or to wavenumbers near $q^{\star}$.  

%The ROL auxiliary field theory and FH theories are both based 
%upon Gaussian approximations for the distribution of fluctuations 
%of either a composition field (in the FH theory) or an auxiliary
%chemical potential field (in the ROL theory) about a saddle point. 
%In the jargon of field theory, this approximation is also referred 
%to as a "one-loop" approximation because it leads to predictions 
%for various physical properties that can represented as diagrams
%that have only a single loop.

%Like the MFEH theory, one-loop auxiliary field theory for $S(q)$ 
%is UV divergent. The treatment of UV divergences presented in Ref. 
%\cite{Piotr_Morse_2007} is based on an explicit demonstration that 
%all UV divergent contributions to the ROL prediction for $S^{-1}(q)$ 
%can be identified physically as contributions to the phenomenological 
%parameters that are required as inputs to the RPA theory. Specifically, 
%it was shown that, for an idealized model with $A$ and $B$ monomers of
%equal statistical segment length, all of the divergent contributions 
%could be identified with renormalizations of either the RPA $\chi$ 
%parameter or the statistical segment length. This allows all of the
%UV divergent contributions to $S^{-1}(q)$ be absorbed into the values 
%of these microscopic parameters, which are then treated as fitting
%parameters when comparing to experiment or simulation.

The ROL theory yields a prediction for $cNS^{-1}(q)$ as the sum of 
an RPA contribution of the form given in eq. (\ref{SInvRPA}), 
with renormalized values for the $\chi$ and $b$ parameters, plus a 
universal correction of the form
\begin{equation}
   cN \delta S^{-1}(q) = 
   \frac{1}{\bar{N}^{1/2}}H(q R,\chi N, f_{A}, b_{A}/b_{B}) .
   \label{deltaSinvScale}
\end{equation}
No simple analytic form exists for the dimensionless function $H$, 
which is evaluated here by numerically integrating a set of related
Fourier integrals (see Sec. III and App. A for details).
%The required integrals and numerical procedure are discussed in 
%Sec. III and App. A. 

We have argued previously \cite{Piotr_Morse_2007, Morse_Qin_2011} 
that the ROL theory appears to be the first correction to the RPA 
in a systematic expansion of an underlying universal expression 
for $cNS^{-1}(q)$ in diblock copolymer melts as a function 
\begin{equation}
  cN S^{-1}(q) = D(q R,\chi N, f_{A}, b_{A}/b_{B}, \bar{N})
  \label{SinvScale}
\end{equation} 
that depends only on the parameters that appear in the RPA and 
on $\bar{N}$. It appears that the ROL theory is part of an expansion of this 
underlying function in powers of $\bar{N}^{-1/2}$, in which an
RPA theory with renormalized parameters is recovered in the limit
$\bar{N} \rightarrow \infty$, and in which the ROL theory is the dominant 
${\cal O}(\bar{N}^{-1/2})$ correction. This claim is based upon a
simple power counting argument \cite{Morse_Qin_2011} that shows that,
if all diagrammatic contributions to $S^{-1}(q)$ within an
infinite series expansion \cite{Morse_2006} are organized into a 
loop expansion, the UV convergent part of the $L$-loop contribution 
will be multiplied by a prefactor of $\bar{N}^{-L/2}$. This 
argument assumes, however, that a renormalization procedure 
similar to what has been used to interpret and remove 
all UV divergent contributions from the one-loop theory for
$S(q)$ can be extended to higher-order contributions to the 
loop expansion. If this is so, the accuracy of the one-loop 
theory should increase systematically with increasing $\bar{N}$, 
for any value of $\chi N$, and for asymmetric as well as symmetric
copolymers. 

We have also discussed the mathematical relationship between the ROL and 
FH predictions for $S(q)$ in more detail elsewhere.\cite{Morse_Qin_2011} 
We showed that, despite differences in how these two theories were 
derived, they yield asymptotically equivalent results for the 
divergence of $\delta S^{-1}(q^{\star})$ for symmetric block 
copolymers near the ODT, where $S(q^{\star})$ is large.
We found that the asymptotic behavior of the ROL prediction for 
$\delta S(q^{*}, \chi N)$ for symmetric copolymers at $q = q_{0}$ 
near the spinodal $\chi N$ is given by eq. (\ref{dSinvFH}), with 
the same value for the numerical prefactor $B$ as that found by
FH. The FH theory is thus a correct asymptotic approximation to 
the full ROL theory for symmetric copolymers, valid for long 
polymers sufficiently near the ODT. Significant differences between 
the predictions of the two theories are thus possible only for
symmetric diblock further from the ODT, and for asymmetric 
copolymers.

%The remainder of this paper is organized as follows: In Section
%\ref{sec:CorrelationFunctions}, we introduce some notation for 
%correlation functions that is needed in what follows, including 
%the Ornstein-Zernike
%equation that we use to distinguish intra- and interm-molecular effects.  
%In Section \ref{sec:ROL} we outline the derivation of the ROL for 
%diblock copolymers that was given in Ref. \cite{Piotr_Morse_2007}.  
%In Section \ref{sec:prediction}, we present the predictions of the 
%ROL for the wavevector and $\chi$ dependence of the correlation 
%function for symmetric and asymmetric
%diblock copolymers, and its description of single-molecule statistics.  In this
%Section, we also compare both the defining equations and predictions of the ROL
%and FH theory, and show that they are asymptotically equivalent close to the
%ODT of symmetric copolymers. A summary of conclusions is given in Section
%\ref{sec:conclusion}.

%-------------------------------------------------------------------------------
\section{Correlation Functions}
\label{sec:CorrelationFunctions}
%-------------------------------------------------------------------------------

We consider a melt of AB diblock copolymers in which each chain 
contains $N$ monomers, or $f_{i}N$ monomers of type $i$, where $i=$ A 
or B.  Let $c = 1/v$ be the overall monomer concentration and $\rho 
= c/N$ be the concentration of molecules. Let $b_{i}$ and $l_{i}=v/b_i^2$ 
denote the statistical segment lengths and packing lengths for monomers of 
type $i$, respectively.  The fluctuating local concentration $c_{i}(\rv)$ 
of $i$ monomers is given by a sum
$c_i(\rv) = \sum_{m,s} \delta(\rv - \Rv_{mi}(s))$, 
in which $\Rv_{mi}(s)$ is the position of monomer $s$ on block $i$ of 
molecule $m$. 

We are primarily interested in the behavior of the structure factor matrix (in
this section, the theory is presented using wavevectors $\qv$ as arguments)
\begin{equation}
  \Sc_{ij}(\qv) 
   = \int d\rv \; \ave{\delta c_i(\rv) \delta c_j(0)} 
                  \ee^{\img\qv\cdot\rv} ,
   \label{SijDef}
\end{equation} 
where $\delta c_i(\rv) =  c_i(\rv) - \ave{c_i}$ is the deviation 
of the monomer concentration $c_{i}(\rv)$ from its mean value
$\ave{c_i} = c f_{i}$. Analogously, we also define an intramolecular correlation
function, $\Ss_{ij}(\qv)$, that arise from correlations between pairs of
monomers on the same chain.

To distinguish the effects of intra- and inter-molecular correlations, it is 
useful to introduce a generalized Ornstein-Zernicke equation
\cite{Erukhimovich_1979, Benoit_1984, Schweizer_1988, Schweizer_1989, Piotr_Morse_2007}
\begin{equation}
  \Sc_{ij}^{-1}(\qv) 
  = \Ss^{-1}_{ij}(\qv) - C_{ij}(\qv),
  \label{OrnsteinZernike}
\end{equation}
in which $\Ss_{ij}(\qv)$ denotes the \textit{true} intramolecular correlation 
function (rather than the random walk approximation used by Leibler), 
and $C_{ij}(\qv)$ is a direct correlation function that is defined by eq. 
(\ref{OrnsteinZernike}).

In a dense, nearly-incompressible liquid with monomers of equal 
volume, we may assume that the eigenmodes of the $2 \times 2$ matrix 
$\Sc_{ij}(\qv)$ for values of $q$ of order the inverse coil size are 
given approximately by a fluctuating composition mode, of the form 
$(\delta c_A(\qv), \delta c_B(\qv)) \propto (1, -1)$, 
and a density fluctuation mode, of the form 
$(\delta c_A(\qv), \delta c_B(\qv)) \propto (1, 1)$. Fluctuations 
of the monomer density mode are strongly suppressed by the low 
compressibility of the liquid, while composition fluctuations can
be quite large. In the limit of negligible density fluctuations, one 
can define a scalar structure function 
\begin{equation}
   \Sc(\qv) = \Sc_{AA}(\qv) = \Sc_{BB}(\qv) 
   = - \Sc_{AB}(\qv) 
   . \label{SscalarDef}
\end{equation}
By combining this assumption of incompressibility with the 
Ornstein-Zernicke equation, it is straightforward to show 
\cite{Schweizer_1988, Schweizer_1989, Piotr_Morse_2007} that $S^{-1}(\qv)$ 
may be expressed in the generalized RPA form
\begin{eqnarray}
  cN S^{-1}(\qv) = F(\qv) - 2\chi_{a}(\qv)N,
  \label{SinvFchiN}
\end{eqnarray}
in which the functions
\begin{eqnarray}
  F(\qv) &\equiv& 
  cN \sum_{ij} \Ss_{ij}^{-1}(\qv) \varepsilon_{i} \varepsilon_{j} ,
  \label{Fdef} \\
  \chi_{a}(\qv) & \equiv & 
  \frac {c}{2} \sum_{ij} C_{ij}(\qv) \varepsilon_{i} \varepsilon_{j} 
  \label{chiadef}
\end{eqnarray}
are defined by projecting $\Ss^{-1}_{ij}(\qv)$ and $-\Ch_{ij}(\qv)$ 
onto the subspace of pure composition fluctuations, where 
$\varepsilon \equiv (\varepsilon_{A}, \epsilon_{B}) = (1, -1)$.
Written more explicitly, this yields expressions
\begin{eqnarray}
  F(\qv) &=& cN \Sssum(\qv) / W(\qv) ,
  \label{FchiaExplicit}\\
  \chi_{a}(\qv) & = & 
  \frac {c}{2} 
  \left [ C_{AA}(\qv) + C_{BB}(\qv) - 2 C_{AB}(\qv) \right ],
\end{eqnarray}
in which 
\begin{eqnarray} 
   \Sssum(\qv) & \equiv & \sum_{ij}\Ss_{ij}(\qv) ,
   \label{intraCorrelationSum}
   \nonumber \\
   W(\qv)    & \equiv & \Ss_{AA}(\qv)\Ss_{BB}(\qv) - \Ss_{AB}^2(\qv)
   .
\end{eqnarray}
We will refer to the wavevector-dependent function $\chi_{a}(\qv)$ 
defined above as the ``apparent'' interaction parameter.

Despite a superficial similarity, eq. (\ref{SinvFchiN}) is 
{\it not} equivalent to the RPA approximation, eq. (\ref{SInvRPA}). 
Eq. (\ref{SinvFchiN}) simply define the
functions $F(q)$ and $\chi_{a}(q)$, by relating these quantities
to the correlations functions $S(q)$ and $\Ss_{ij}(q)$, without
assuming anything about the behavior of these functions. The 
RPA is instead obtained by approximating the intramolecular 
correlation functions used to define $F(q)$ by those of a
Gaussian chain, and approximating $\chi_{a}(q)$ by a parameter 
$\chi$ that is independent of both wavenumber $q$ and chain 
length $N$.
%The function $F(\qv)$ in eq.  (\ref{SinvFchiN}) is defined using the 
%true intramolecular correlation functions $\Ss_{ij}(\qv)$, while 
%the function $F_{0}(\qv)$ in eq. (\ref{SinvFchiN}) is calculated by 
%using a Gaussian random-walk model for these functions.
%Eq. (\ref{chiadef}) simply defines $\chi_{a}(\qv)$, but 
%makes no assumptions about how it depends on $\qv$ or chain length,
%while eq. (\ref{SInvRPA}) assumes that $\chi_{a}(\qv)$ 
%can be approximated by a parameter $\chi$ that is independent 
%of wavenumber $q$ and chain length $N$. 
The most general form of eq. (\ref{SInvRPA}) can allow $\chi$
to exhibit an arbitrary dependence upon both temperature and 
composition, but would cease to have any predictive value if 
it also allowed for an arbitrary dependence on $\qv$ and $N$. 
%Eq. (\ref{SinvFchiN}) is thus a complete general expression, 
%which allows for corrections to the RPA that arise both from
%corrections to the random-walk model for intramolecular 
%correlations and corrections to the SCF treatment of 
%interactions.

%The random walk model that underlies the RPA yields intramolecular 
%corrrelation functions
%%
%\begin{eqnarray}
%  \Ssi_{ii}(\qv) & = & \rho N^2 f_i^2 g(q^2 R^2_{g,i})
%  \nonumber \\
%  \Ssi_{ij}(\qv) & = &
%  \rho N^2 f_i f_j e(q^2 R^2_{g,i}) e(q^2R^2_{g,j}),
%\end{eqnarray}
%for $j \neq i$,
%where $R^2_{g,i} = f_i N b_i^2/6$ is the radius of gyration of 
%the $i$ block in a diblock copolymer with $f_{i}N$ monomers of 
%type $i$, and where
%\begin{eqnarray}
%   g(x) & \equiv & 2(e^{-x}-1+x)/x^2 \nonumber \\
%   e(x) & \equiv & (1 - e^{-x}) / x
%\end{eqnarray}
%Here and hereafter, we use a tilde $\tilde{\ }$ to identify 
%properties of ideal Gaussian chains. 

%-------------------------------------
\section{Renormalized One-Loop Theory}
\label{sec:ROL}
%-------------------------------------

In this section, we briefly review the main results of the ROL theory
for $S(\qv)$ in diblock copolymer melts.\cite{Piotr_Morse_2007}

The derivation of the ROL theory is based on a coarse-grained model 
in which the total potential energy $U$ is the sum of intramolecular 
potential $U_\text{chain}$ of a set of Gaussian chains plus a 
short-range non-bonded pair interaction.  We consider a pair potential 
of the form
%\begin{equation}
$U_{ij}(\rv) = v_{0}\epsilon_{ij}\delta_{\Lambda}(\rv)$,
%\end{equation}
where $\delta_{\Lambda}(\rv)$ denotes a short-range function with 
a unit integral, $\int d\rv \delta_\Lambda(\rv) = 1$, and with a 
characteristic range of interaction $\Lambda^{-1}$. 
The interaction parameters $\epsilon_{ij}$ are given by
$\epsilon_{AA} = \epsilon_{BB} = B_{0}$ and
$\epsilon_{AB}  =  B_{0} + \chi_{0}$,
where $B_{0}$ is a dimensionless compression modulus, and where 
$\chi_{0}$ is a ``bare'' Flory-Huggins interaction parameter.

The one-loop approximation for $\Sc_{ij}(\qv)$ given in ref. 
\cite{Piotr_Morse_2007} was based on a more general diagrammatic 
expansion that was discussed in ref.  \cite{Morse_2006}. There, 
the details were shown on the construction of expansions of various correlation 
functions in terms of cluster diagrams that represent Gaussian 
chains interacting via a screened interaction $\Gh$. The Fourier
transform $\Gh(\kv)$ of this screened interaction is given by
\begin{equation}
   \Gh^{-1}_{ij}(\kv) = \Ssi(\kv) + U^{-1}_{ij}(\kv) ,
   \label{Gdef}
\end{equation}
in which $\Gh_{ij}(\kv)$, $U_{ij}(\kv)$ and $\Ssi_{ij}(\kv)$ are 
$2 \times 2$ matrices, and inversion refers to matrix inversion.
This is a straightforward generalization to multi-component
systems of the screened interaction introduced by Edwards to 
describe excluded volume interactions in concentrated homopolymer 
solutions. The limit of incompressible dense liquids is 
obtained in this formalism by taking $B_{0} \rightarrow \infty$. 

%In the limit of incompressible 
%two-component polymers (i.e., blends or diblock copolymer melts)
%this expression for $G_{ij}(\qv)$ reduces to \cite{Piotr_Morse_2007}
%\begin{equation}
%  \Gh_{ij}(\qv) = \frac
%    { 1 - 2 v \chi_0 |\Ssi(\qv)| \Ssi^{-1}_{ij}(\qv) }
%    { \Sssum(\qv) - 2 \chi_0 v |\Ssi(\qv)| }
%    \label{screenedInteraction}.
%\end{equation}
%for $q \ll \Lambda$.

The one-loop approximation for $\Sc(\qv)$ is based upon an analysis 
of the two diagrams shown in Fig. 1. The diagram on the left yields 
the one-loop contribution to the intramolecular correlation function 
$\Ss_{ij}(\qv)$, while the one on the right yields a one-loop 
contribution to the direct correlation function $C_{ij}(\qv)$. 

\subsection{Intramolecular Correlations}
\begin{figure}[htb]
  \centering
  \includegraphics[width=.35\textwidth,height=!]{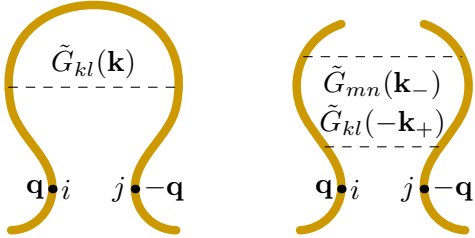}
  \caption[Illustrative diagrams for intra- and intermolecular correlations.]
  {Diagrams illustrating the effects of the screened intra- (left) and
   inter- (right) molecular interactions on the correlations of segments 
   of type $i$ and $j$. Additional segmental indices $k$, $l$, $m$, $n$ 
   runs through the whole chain. $\kv$ and $\qv$ are wave vectors for the 
   interactions or correlations, and $\kv_\pm \equiv \kv \pm \frac{\qv}{2}$ 
   is one particular choice made to conserve momentum.}
   \label{oneloopDiagram}
\end{figure}

The one-loop theory predicts an intramolecular correlation function
\begin{equation}
    \Ss_{ij}(\qv) = \Ssi_{ij}(\qv) + \Delta \Ss_{ij}(\qv) ,
\end{equation}
in which $\Ssi_{ij}(\qv)$ is the correlation function for a gas
of noninteracting Gaussian chains. The correction $\Delta \Ss_{ij}(\qv)$ 
arising from interactions is given by a Fourier integral
\begin{equation}
  \Delta \Ss_{ij}(\qv) = - \frac{\rho}{2} \int_\kv
             \psi_{ijkl}^{(4)}(\qv,-\qv,\kv,-\kv) \Gh_{kl}(\kv).
 \label{intradef}
\end{equation}
Here and hereafter, summation over repeated subscripts is 
implicit, and $\int_\kv \equiv (2\pi)^{-3} \int d\kv$.  The four-point 
function $\psi_{ijkl}^{(4)}(\qv,-\qv, \kv, -\kv)$ is defined by
\begin{equation}
  \psi_{ijkl} = \ssi^{(4)}_{ijkl}(\qv,-\qv,\kv,-\kv)
              - \ssi_{ij}(\qv) \ssi_{kl}(\kv),
\end{equation}
in which $\ssi_{kl}(\kv) \equiv \Ssi_{kl}(\kv)/\rho$ is a single-chain
correlation function, normalized by the molecular density $\rho = c/N$, 
and $\ssi^{(4)}_{ijkl}(\qv, -\qv, \kv, -\kv)$ is an analogous single-chain 
four point function for monomers within blocks $i$, $j$, $k$, and
$l$.\cite{Piotr_Morse_2007} The physical content of eq. (\ref{intradef}) 
is shown schematically in the left diagram of Fig. \ref{oneloopDiagram}, 
in which the curve represents a single chain, along which two segments 
interact via a screened potential $\Gh$.

The integral in eq. (\ref{intradef}) is ultraviolet (UV) divergent: 
The value of the Fourier integral in this equation is dominated by 
contributions of large wavevectors, and diverges in the absence of 
a high-$\kv$ cutoff. If the range of $\kv$ is restricted to
$|\kv| < \Lambda$, the value of the integral increases linearly with 
the cutoff wavenumber $\Lambda$ for $\Lambda R_{g} \gg 1$.  

It was shown in the previous work,\cite{Piotr_Morse_2007} however, 
that this type of sensitivity of the one-loop prediction for $\Ss_{ij}(\kv)$ 
to short-wavelength correlations can be entirely attributed to 
the change in the value of the effective statistical segment 
length. Specifically, it was shown that the ROL result for 
$\Ss_{ij}(\qv; \Lambda)$ with a cutoff wavenumber $\Lambda$ is 
consistent with an expression of the form 
\begin{equation}
   \Ss_{ij}(\qv) = \Ssi_{ij}(\kv; b(\Lambda))
   + \delta \Ss_{ij}(\qv),
\end{equation}
in which $\Ssi_{ij}(\kv; b)$ is the correlation function for 
a Gaussian chain with a cutoff-dependent statistical segment 
length $b(\Lambda)$ that is different than that of a polymer
in vacuum, as the result of interactions between polymers in a 
dense liquid, and in which $\delta \Ss_{ij}(\qv)$ is a small
cutoff-independent correction to Gaussian statistics. The 
treatment of intramolecular correlations used here is a 
generalization of the results of Wittmer, Beckrich {\it et al.},
\cite{Wittmer_2004, Beckrich_Wittmer_2007,Wittmer_Beckrich_2007} 
who used a equivalent form of ROL theory to successfully 
predict universal corrections to Gaussian chain statistics 
for homopolymers in a dense melt. In a diblock copolymer 
melt, the predicted correction to the random-walk model for 
$\Ss_{ij}(\qv)$ is given by a function of the form
\begin{equation}
  \delta \Ss_{ij}(\qv) = \frac{cN}{\bar{N}^{1/2}}
         \delta \hat{\ss}_{ij}(\qv R_g, \chi_0 N, f_A, b_B/b_A),
\end{equation}
and is smaller than the ideal gas contribution $\Ssi_{ij}(\qv)$
by a prefactor of $\bar{N}^{-1/2}$. Here, $\hat{\ss}_{ij}$ is 
a dimensionless function that, in the case of interest here, 
can only be evaluated numerically, by a procedure outlined in 
App. \ref{appdx:renormalization}. The corresponding correction to the random walk model 
for $F(\qv)$ is given by
\begin{equation}
   \delta F(\qv) = - cN \sum_{ij}\varepsilon_i
   \Ssi^{-1}_{ij}(\qv) \delta\Ss_{jk}(\qv)
   \Ssi^{-1}_{kl}(\qv) \varepsilon_l .
\end{equation}
This is the expression that we use to evaluate the intramolecular
one-loop correction to $S^{-1}(\qv)$.

\subsection{Direct Correlation Function}
The one-loop theory yields an expression for the apparent 
$\chi$-parameter as a sum
\begin{equation}
   \chi_{a}(\qv) = \chi_{0} + \Delta \chi_{a}(\qv),
\end{equation}
in which
\begin{eqnarray}
  \Delta \chi_{a}(\qv) & = &
  c\sum_{ij}\varepsilon_{i}\varepsilon_{j}
  \Ssi_{ik}^{-1}(\qv) \Sigma_{kl}(\qv) \Ssi_{lj}^{-1}(\qv)
  ,
  \label{interdef} \\
  \Sigma_{ij}(\qv) &=& 
  \frac{1}{2} \int_\kv \Ssi^{(3)}_{imk}(\qv,\kv_-,-\kv_+)
  \Gh_{kl}(\kv_+) \nonumber\\
  & &\ \times \Ssi^{(3)}_{jnl}(-\qv,-\kv_-,\kv_+) \Gh_{mn}(\kv_-)
  . 
  \label{sigmadef}
\end{eqnarray}
Here, $\kv_\pm \equiv \kv \pm \frac{\qv}{2}$, and $\Ssi^{(3)}_{ijk}$ is a
three point intramolecular correlation function for monomers within
blocks $i$, $j$ and $k$ of a Gaussian diblock. The physical content 
of eq. (\ref{sigmadef}) for $\Sigma_{ij}$ is shown schematically by 
the right diagram of Fig. \ref{oneloopDiagram}, in which two pairs 
of segments along two chains interact via the screened interaction 
$\Gh$.  

The integral in eq. (\ref{sigmadef}) for $\Sigma_{ij}(\qv)$ is also 
UV divergent, and so yields a UV divergent expressions for $\Delta 
\chi_{a}(\qv)$. It was shown in ref. \cite{Piotr_Morse_2007}, 
however, that the divergence of this quantity could be interpreted
as a renormalization of the interaction parameter. It was shown that,
in a model with equal A and B statistical segment lengths, 
the one-loop prediction for $\chi_{a}(\qv)$ may be written as a 
sum of the form
\begin{equation}
   \chi_{a}(\qv) = \chi_{e}(\Lambda) + \delta \chi_{a}(\qv)
   ,
\end{equation}
in which $\chi_{e}(\Lambda)$ is a cutoff-dependent effective
interaction parameter that is independent of both $\qv$ and $N$,
and $\delta \chi_{a}(\qv)$ is a cutoff-independent correction. 
The correction $\delta \chi_{a}(\qv)$, which depends upon 
both $\qv$ and $N$, represents a correction to the phenomenology 
predicted by the RPA, rather than merely a correction to the 
value of the interaction parameter. It was found that this 
quantity is given by a function of the form
\begin{equation}
  \delta \chi_{a} (\qv) = \frac{1}{N \bar{N}^{1/2}} 
  \delta \hat{\chi}_{a} (\qv R_g, \chi N, f_A, b_B/b_A),
\end{equation}
where $\delta \hat{\chi}_{a}$ is a dimensionless function that
we calculate by numerical integration. 

%Later on, all results are reported with respect to $\chi_e$ in 
%favor of $\chi_0$. When Eq. (\ref{intradef}) and (\ref{interdef}) 
%are used to evaluate corrections, the $\chi_0$ terms are also 
%replaced with $\chi_e$, with the understanding that the difference 
%is irrelevant at the one-loop order. Similar to 
%$\delta \Ss_{ij}^\star(\qv)$, $\delta\chi^\star(\qv)$ becomes significant 
%near the mean field spinodal.

%-----------------------------------------
\subsection{Self-Consistent Approximation}
\label{sec:SCA}
%-----------------------------------------

The FH theory for $S(\qv)$ in the disordered phase is based on a 
self-consistent one-loop, or Hartree, treatment of fluctuations. Such 
an approximation is obtained by starting from a perturbative one-loop 
approximation for $\delta S(\qv)$, in which $\delta S(\qv)$ is initially
expressed as a Fourier integral involving the RPA or ``bare'' correlation 
function $S_{0}(\qv)$, and then replacing $S_{0}(\qv)$ by the self-consistently
determined correlation function $S(\qv)$ throughout the expression for 
$\delta S(\qv)$. The use of this approximation is justified for symmetric 
diblock copolymers near the ODT by an analysis given by Brazovski\u\i,
who concluded that the Hartree approximation correctly captures the 
dominant contributions to $\delta S(\qv)$ near the spinodal of any
homogeneous systems for which Landau theory predicts a second order 
transition to a periodic state.

To define an analogous self-consistent approximation for the auxiliary 
field theory, we simply evaluate all of our expressions for the one-loop 
correction $\delta \Omega_{ij}(\qv R, \chi N)$ and $\delta F(\qv R, \chi N)$ 
by replacing $\chi$ by an ``apparent" $\chi$-parameter, $\barchia$ that is 
defined by fitting the actual peak-intensity to the RPA, by setting
\begin{equation}
   c N S^{-1}(q^{\star}) = 2[ (\chi N)_{s} - \barchia N ]
   , \label{chibardef}
\end{equation}
where $(\chi N)_{s}$ denotes the RPA spinodal value. We thus evaluate 
the one-loop corrections using a value of $\chi$ that is chosen so that 
the underlying approximation for $S(\qv)$ has the correct, self-consistently
determined peak value. In this approximation, we have
\begin{align}
   S^{-1}(\qv) = S^{-1}_{0}(\qv) + \delta S^{-1}(\qv R, \barchia N)  .
\end{align}
Here, the function $\delta S^{-1}(\qv R, \barchia N)$ represents the one-loop 
correction obtained from the perturbative one-loop theory, after subtracting
UV divergent terms. In the same approximation, the microscopic parameter 
$\chi_{e}$ (which is assumed to depend on temperature and details of local 
liquid structures, but not on $N$) is related to $\barchia$ by a 
relationship
\begin{equation}
  \barchia N = \chi_e N 
     - \frac{c N}{2} \delta S^{-1}(q^\star R, \barchia N) .  
  \label{chi_selfconsistent}
\end{equation}
The self-consistent theory can be evaluated numerically by calculating the 
one-loop contribution for a specified sequence of values of $\barchia N$, 
and then using eq. (\ref{chi_selfconsistent}) to infer corresponding 
values of $\chi_e N$.

%---------------------
\section{Results}
\label{sec:prediction}
%---------------------

We present the predictions of the ROL theory in this section, and compare 
them to the RAP and FH theories whenever possible. Since we are interested 
in the disordered phase throughout this work, from now on, we use wavenumbers
$q=|\qv|$ instead of vectors $\qv$ as arguments.

%---------------------------------------------
\subsection{``Ideal'' Symmetric Copolymers ($\chi = 0$)}
\label{subsec:idealcopolymer}
%---------------------------------------------

We begin by considering the special case of a model with 
$\chi_{0}=0$, in which the underlying pair interaction $U_{ij}(\rv)$ 
is independent of the monomer type indices $i$ and $j$. This 
describes a homopolymer melt in which we arbitrarily label the 
first $f_{A}N$ monomers of each chain as A monomers, and the 
remainder as B, but in which A and B monomers are 
physically indistinguishable. This situation is approximately 
realized in neutron scattering experiments in which the two 
blocks of a diblock molecule contain protonated and deuterated versions 
of the same monomer. We refer to this as an ``ideal" copolymer 
by analogy to the description of a mixture of two almost
indistinguishable molecular species as an ``ideal'' solution. 

We further focus our attention on the case of symmetric diblock 
copolymers, with $f_{A}=f_{B}=1/2$ and $b_{A}=b_{B}$, and in 
which the two blocks are also structurally identical. We show 
in App. \ref{app:zeroChi} that, in this case, $\chi_{a}(q)=0$ 
for all $q$ when $\chi_{0}=0$. The inverse correlation 
function of an incompressible melt of such polymers thus is
given exactly by
\begin{equation}
   cN S^{-1}(q) = F(q) ,
\end{equation}
where $F(q)$ is the combination of intramolecular correlation
functions defined in eq. (\ref{Fdef}). In this case, the 
deviation of $S^{-1}(q)$ from Leibler's RPA prediction for
ideal diblocks is thus entirely a result of a deviation from
Gaussian single-chain statistics. 
 
\begin{figure}[tb]
 \centering
 \subfigure[]{\includegraphics[width=.4\textwidth,height=!]{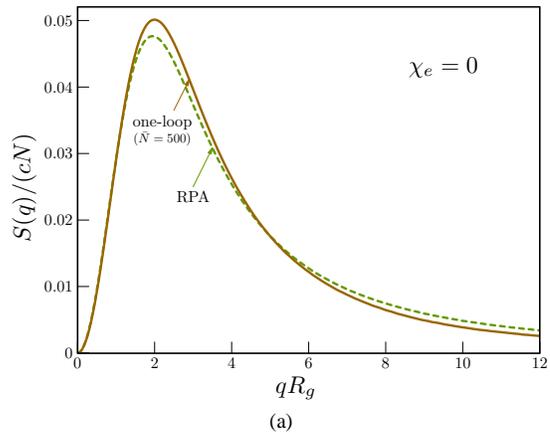}\label{sofq}}
 %\subfigure[]{\includegraphics[width=.4\textwidth,height=!]{deltaF}\label{deltaF}}
  \caption[Wavevector dependence of correlation function for diblocks at
   $\chi_e = 0$ for $\bar{N}=500$.]
  {Normalized correlation function $cN S(q)$ for symmetric diblock 
   copolymer with $\chi_{e} = 0$ and $\bar{N}=500$, as predicted by 
   the RPA (dashed line) and ROL (solid) theories. }
\end{figure}
 
Fig. \ref{sofq} shows predictions for the normalized structure factor 
$S(q)/cN$ for a melt of symmetric diblock copolymers with $\bar N = 500$ 
and $\chi_e = 0$, as predicted by original RPA theory (dashed line) and 
by the ROL theory (solid line).  The one-loop corrections to the RPA
cause a slight increase in peak position $q^{\star}$, to a value 
slightly above the RPA value $q_{0}$, and cause a slight enhancement in 
peak intensity $S(q^\star)$. Both of these trends are opposite to those 
found near the ODT, where composition fluctuations tend to decrease 
$q^{\star}$ and suppress the peak intensity. For large $q$ values, 
$qR_{g} > 5$, the correction to the RPA becomes negative, and is found 
to decrease asymptotically as $q^{-3}$ for $qR_{g} \gg 1$. The behavior 
in this high-q regime is identical to that found by the Strasbourg group 
\cite{Beckrich_Wittmer_2007, Wittmer_Beckrich_2007} for the high-$q$ 
corrections to the intramolecular correlation function in a homopolymer 
melt. 

\begin{figure}[tb]
 \centering
 \includegraphics[width=.4\textwidth,height=!]{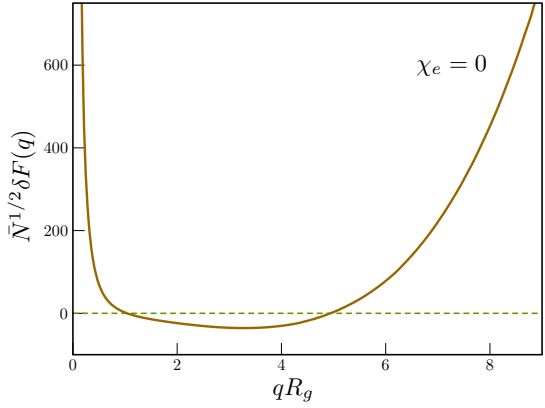}\label{deltaF}
 \caption[Wavevector dependence of $\delta F(q)$ for symmetric diblocks 
 with $\chi_e = 0$ and $\bar{N}=500$.]
  {Normalized one-loop correction $\bar{N}^{1/2} \delta F(q)$ to
   the inverse correlation function for symmetric diblock copolymers 
   with $\chi_{e} = 0$.}
\end{figure}

The predicted corrections to the RPA are proportional to $\bar{N}^{-1/2}$,
and so would be smaller than shown in Fig. \ref{sofq}) for longer chains.
In Fig. \ref{deltaF}, we show the one-loop correction $\delta F(q)$ to 
the random walk model for $F(q)$, multiplied by $\bar{N}^{1/2}$ to obtain 
a universal function of $qR_{g}$ that is independent of $\bar{N}$. 
Three different wavenumber regimes are visible in this representation:
For small wavenumbers, $qR_{g} < 1$, both $F_{0}(q)$ and $\delta F(q)$ 
are positive, and both diverge as $1/q^{2}$ as $q \rightarrow 0$. We 
show in Sec. IV that $F(q) \propto 1/(qR_{AB})^{2}$ for $q R_{g} \ll 1$, 
where $R_{AB}^{2}$ is the mean-squared separation between the centers 
of mass of the A and B blocks. A positive correction to $F(q)$ in this 
limit thus corresponds to a negative correction to the random walk 
prediction for $R_{AB}$. At very high wavenumbers, $qR_{g} \gg 5$, 
we find $\delta F(q) \sim  \bar{N}^{-1/2}(qR_{g})^{3}$ and 
$F_{0}(q) \sim (q^{2}R_{g})^{2}$, which leads to the behavior 
$cN\delta S(q) \sim -\bar{N}^{-1/2}(q R_{g})^{-3}$ noted above.
In the intermediate range $1 < qR_{g} < 5$ surrounding $q_{0}R_{g} 
\simeq 2$, however $\delta F(q)$ is negative and decreases with
increasing $q$, leading to an enhancement in $S(q)$ and a slight 
positive shift in $q^{*}$. 

The ROL theory predictions of a value of $q^{*}$ slightly higher than the
RPA prediction, and of a value of $R_{AB}$ slightly less than the 
random walk prediction, are meaningful only if accompanied by a
clear definition for the value of the statistical segment length 
$b$ used in the random-walk model to which the ROL theory is compared. 
The way the ROL theory is derived requires that the statistical 
segment length in the RPA theory be defined to be that of a 
hypothetical system of infinite chains, by defining $b^{2}$ to
be the $N \rightarrow \infty$ limit of the ratio $6R_{g}^{2}/N$ 
in a homologous series of melts of increasing chain length $N$. 
This is the definition used by the Strasbourg group to compare 
their ROL predictions for the intramolecular correlation function 
in homopolymer melts to the results of extensive computer simulations.
\cite{Wittmer_2004, Beckrich_Wittmer_2007,Wittmer_Beckrich_2007}

%----------------------------------------------------
\subsection{Peak Intensity (Symmetric Copolymers, $\chi \neq 0$)}
\label{subsec:PeakIntensity}
%----------------------------------------------------

\begin{figure}[tb]\centering
  \subfigure[]{\includegraphics[width=.37\textwidth,height=!]{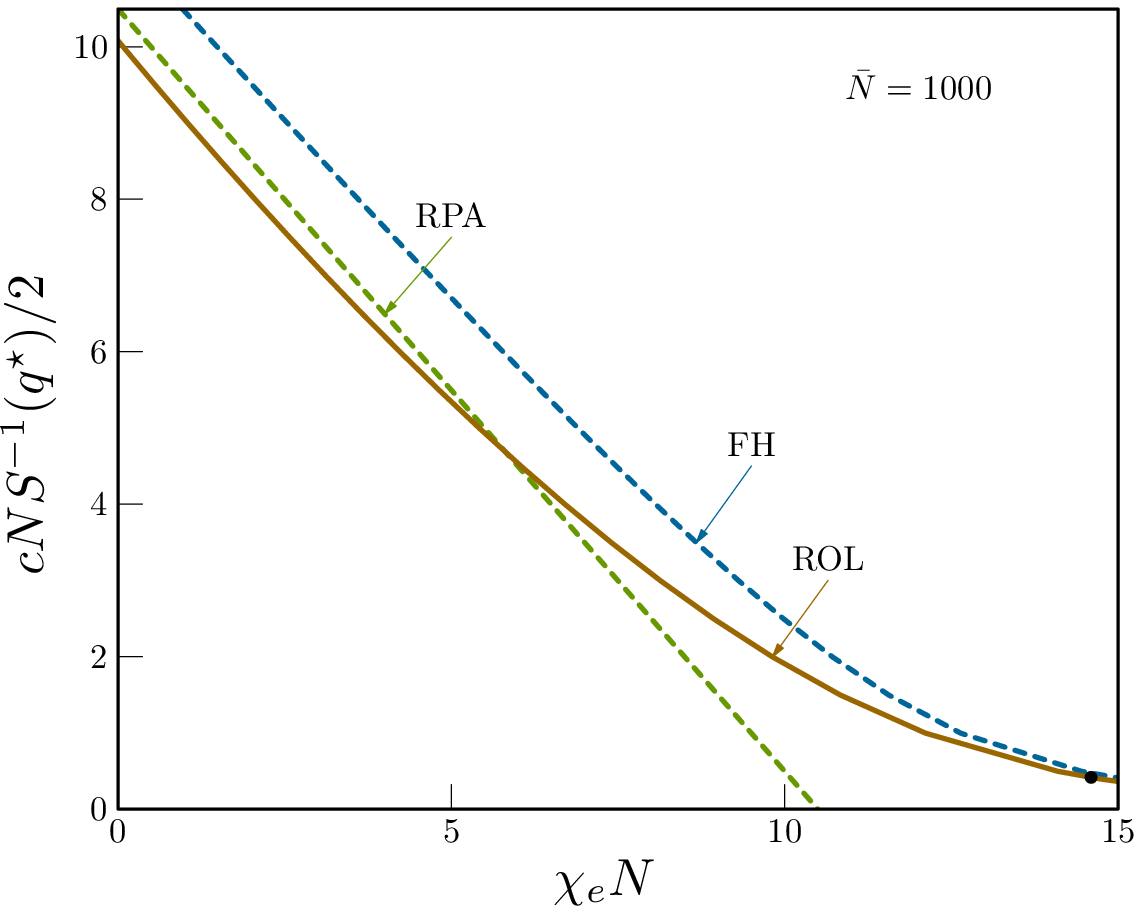}}
  \subfigure[]{\includegraphics[width=.37\textwidth,height=!]{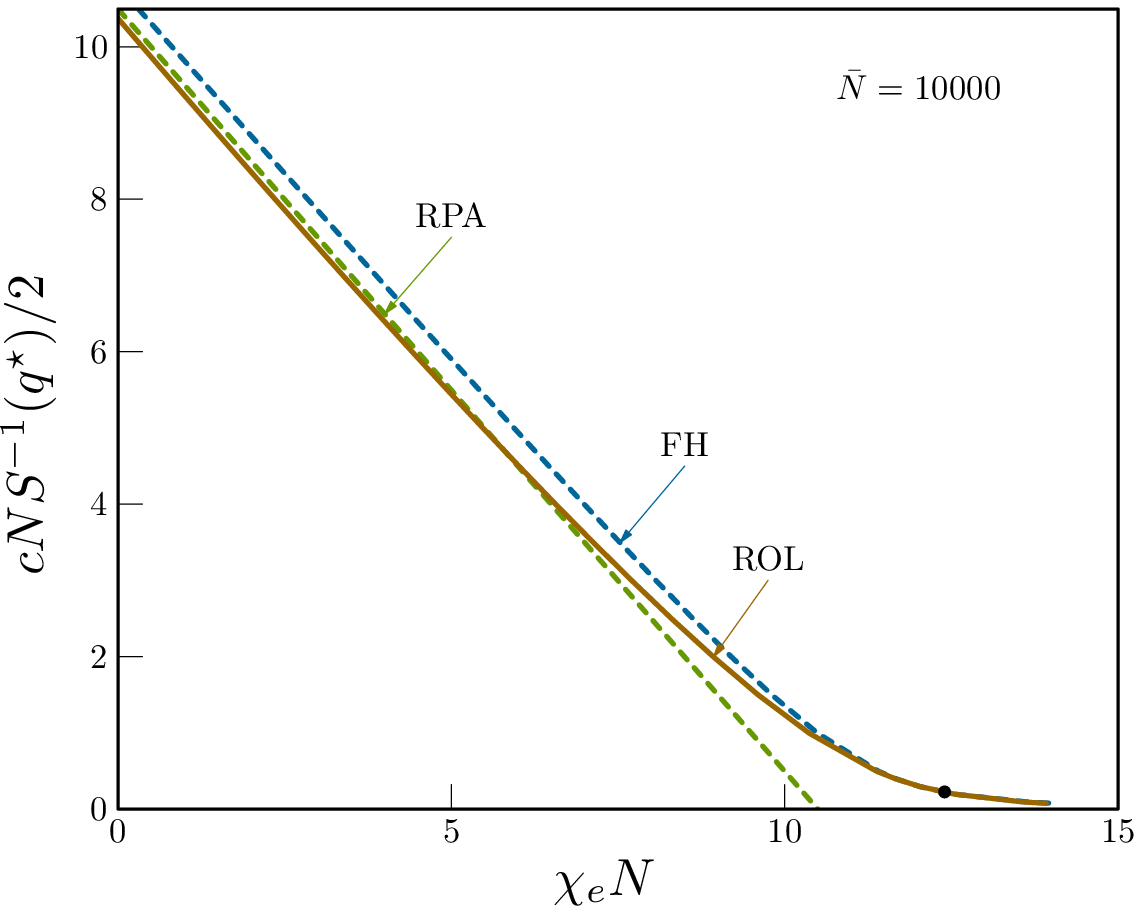}}
  \caption[Self-consistently calculated peak intensity.]
  {Self-consistently calculated inverse peak intensity versus $\chi_e N$
   for symmetric diblock copolymer with $\bar N = 1000$ (a) and 10000 (b).
   Predictions of RPA, the FH theory, and the ROL theories are shown.
   The dots marks the ODTs predicted using the FH theory.}
  \label{scfsinv}
\end{figure}

We now consider how the value of the maximum $S(q^{\star})$ in the 
structure factor for a symmetric diblock copolymer evolves with increasing
$\chi$.  Fig. \ref{scfsinv} compares predictions of the RPA, FH, and ROL 
theories for the normalized inverse intensity $cN S^{-1}(q^\star)$ vs. 
$\chi_e N$ for symmetric diblock copolymers with $\bar N =$ 1000 and 
10000. This corresponds roughly to the range explored by most 
experiments.  The straight dashed line is the RPA prediction, in which 
$S^{-1}(q^\star)$ is a linear function that vanishes (or $S(q^{\star})$
diverges) at $\chi_e N = 10.495$.  Both the FH and ROL theories predict 
deviations from the RPA that decrease as $\bar{N}^{-1/2}$ with 
increasing chain length, but that remain significant at experimentally 
relevant chain lengths. The dots are the FH theory predictions for the 
values of $\chi N$ at the fluctuation-induced first-order transition. 
The ROL auxiliary field theory does not yet provide a prediction for 
the this transition.

The FH theory and the self-consistent ROL theory predict similar results 
at large values of $\chi_e N$, where the deviations from the RPA are 
dominated by the effects of strong composition fluctuations with 
wavenumbers $q \simeq q^{\star}$. This is the regime that the FH theory 
was designed to describe, and in which it has been most successful as a 
description of experimentally measured scattering intensities in nearly 
symmetric diblock copolymers.
\cite{Bates_Rosedale_1988, Rosedale_Bates_1995, Almdal_Bates_1996,
Sakamoto_Hashimoto_1995, Bartels_Stamm_1994, Owens_Russel_1989, Wolff_Ruland_1993} 
Note that deviations from the RPA predicted by both FH and ROL theories 
remain significant even far from the predicted ODT, particularly for the FH 
theory at $\bar{N} = 1000$, that the differences between the FH and ROL 
predictions increase as $\chi_e N$ decreases. The FH theory always predicts 
a positive correction to the RPA expression for $S^{-1}(q^{\star})$, or a 
decrease of $S(q^{\star})$. The ROL theory predicts a positive correction 
to $S^{-1}(q)$ (a suppression of scattering) for $\chi_e N \gtrsim 6$ but 
a negative correction (an enhancement of $S(q^{\star})$) at smaller values 
of $\chi_e N$.  This allows the ROL predictions to interpolate smoothly 
between the behavior found at $\chi = 0$, where the theory yields an 
enhancement of $S(q^{\star})$, and behavior similar to that of the FH theory 
near the ODT. 

\begin{figure}[tb]
\centering
\includegraphics[width=.4\textwidth,height=!]{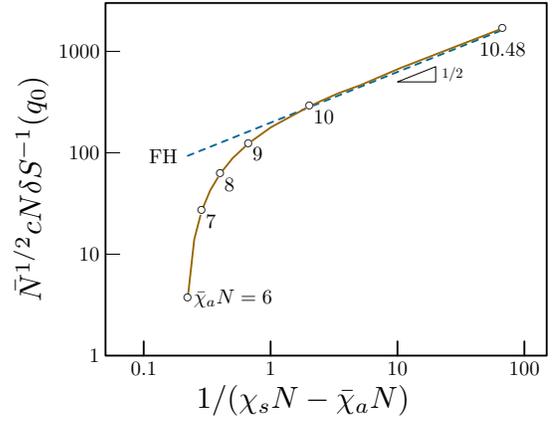}
\caption[Asymptotic comparison between Fredrickson-Helfand theory and ROL theory.]
{Correction to $S^{-1}(q_0)$ predicted by the FH theory (dashed) and by the ROL
 theory (solid) on log-log scale. The
 two theories exhibit the same dominant asymptotic behavior as $\barchia$ 
 approaches $\chi_s$.}
\label{asymptotics}
\end{figure}

Fig. \ref{asymptotics} shows a more careful comparison of the 
asymptotic behavior of the ROL in the limit of strong scattering. 
There, we show a log-log plot of the predicted one-loop correction 
$\bar{N}^{1/2}c N\delta S^{-1}(q_{0})$ as a function of 
$1/[(\chi N)_{s} - \barchia N]$. The square-root divergence predicted 
by the FH theory, given by eq. (\ref{dSinvFH}), is shown by the straight 
line. The convergence of ROL towards this line confirms the conclusion 
of ref. \cite{Morse_Qin_2011} that the FH and the self-consistent ROL 
exhibit the same asymptotic behavior near the spinodal. The differences 
between the two theories become significant for 
$\barchia N \alt 9.5$.

The first, and most influential, quantitative comparison between the FH theory 
and experiments was given by Bates {\it et al.},\cite{Bates_Rosedale_1990} 
for melts of relatively high molecular weight nearly symmetric ($f_A = 0.55$) 
poly(ethyl-propylene-b-ethyl ethylene) (PEP-PEE) diblock copolymer. These
authors concluded that the FH theory gave a quantitatively accurate description 
of both the temperature dependence of the peak intensity near the ODT for 
systems of fixed molecular weights, and of the relationship between the ODT
temperatures of several samples of different molecular weights, using a single 
function $\chi(T) = A/T + B$. In the original study, $\bar{N}$ was estimated 
by fitting $q^{\star}(T)$ to the RPA prediction, thus assuming that $q^{\star}
R_{g}$ was constant. This gave a value of $\bar{N}$ that varied over the
range $6000 < \bar{N} < 11000$ for the most heavily studied sample over the 
experimental temperature range, which was found to correspond to a range
$10.6 < \chi N < 12.5$.  Using data reported in a later work,
\cite{Rosedale_Bates_1995} we have also estimated $\bar{N}$ by using 
independently determined statistical segment lengths for homopolymers. This 
procedure yields noticeably lower values of $\bar{N}$: $3400 < \bar{N} < 4300$. 
This difference suggests that the near perfect agreement between the FH theory
and the experimental results for both the scattering in the disordered phase
and the ODT may be partly fortuitous.  Upon comparing predictions of the FH 
and ROL theories in the relevant range of values of $\chi N$, however, using
either method of estimating $\bar{N}$, we find that these theories yield very 
similar results for such high values of $\bar{N}$ over the limited range of 
values of $\chi N$ probed in these experiments. We thus doubt that it is 
possible to distinguish between the two theories by comparing only their 
predictions for the disordered state scattering intensities reported in 
this study. Quantitative comparisons to lower molecular weight samples over 
a wider range of values of $\chi N$ would highlight the differences, as 
would comparison to computer simulations.

%---------------------------
\subsection{Peak Wavenumber (Symmetric Copolymers, $\chi \neq 0$)}
%---------------------------

\begin{figure}[tb]
  \centering
  \subfigure[]{\includegraphics[width=.22\textwidth,height=!]{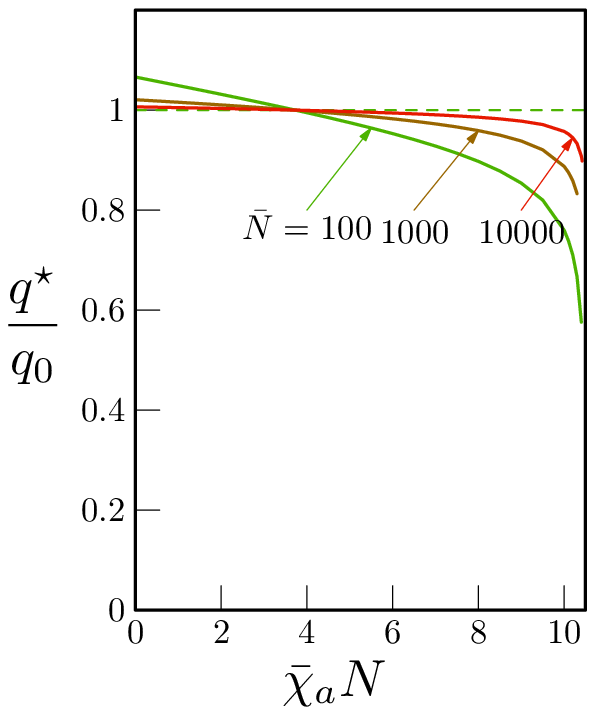}}
  \subfigure[]{\includegraphics[width=.225\textwidth,height=!]{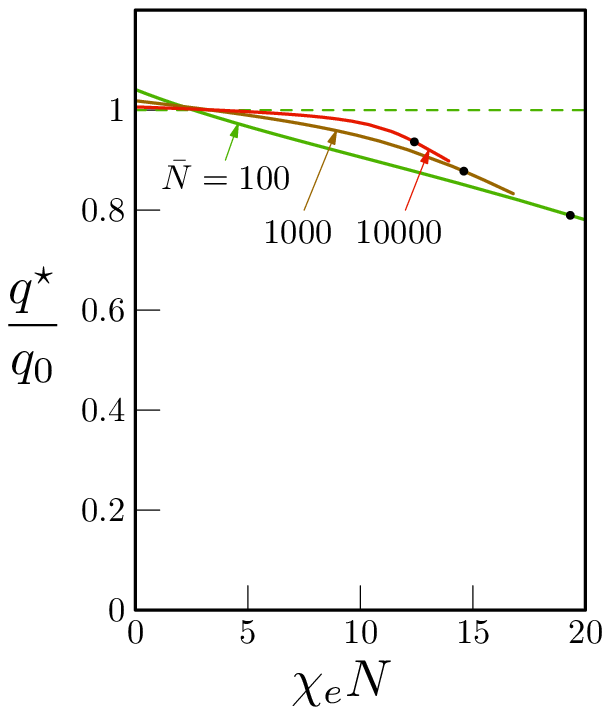}}
  \caption[One-loop theory predictions of the shifts in $q^\star$.]
  {Shift in the peak position of correlation function versus
   (a) $\barchia N$ (b) and $\chi_e N$ for symmetric diblocks with $\bar N
   = 100$, 1000 and 10000, respectively. The sub-figure (b) is prepared using the
   self-consistent approximation discussed in Sec. \ref{sec:SCA}. The dots in
   (b) are the FH theory predictions for $(\chi N)_{ODT}$.}
   \label{qvschiN}
\end{figure}

The ROL theory predicts a peak wavenumber $q^\star$ for symmetric diblock copolymers 
that decreases slightly with increasing $\chi_e N$, with a more rapid decrease near 
the transition.  Fig. \ref{qvschiN} shows predictions of the normalized peak 
position $q^{\star}/q_{0}$ vs. $\chi N$ for several different chain lengths, where 
$q_{0} = 1.946/R_{g}$ is the RPA prediction.\cite{Leibler_1980}

Sub-figure (a) shows the peak wavenumber vs. $\barchia N$. Since the definition
of $\barchia$ in eq. (\ref{chibardef}) is directly related to the peak
intensity, this is essentially a plot of peak wavenumber versus a measure of
peak intensity. Sub-figure (b) instead shows the peak position vs.  $\chi_{e}N$. 
Because $\chi_{e}$ is expected to be a function of temperature alone, and nearly 
linear in $1/T$, sub-figure (b) shows the predicted behavior of an experimental 
plot of $q^{\star}$ vs. $N$ for a sequence of polymers at the same temperature, 
or of $q^{\star}$ vs. $1/T$ in a system for which the statistical segment lengths 
have negligible intrinsic temperature dependence (or for which the data has been
plotted in a way that corrects for this effect).

The predicted value of $q^{\star}$ is slightly larger than $q_{0}$ at $\chi_{e} N = 0$, 
as already noted in Sec. \ref{subsec:idealcopolymer}, and less than $q_{0}$ near 
the ODT.  The dots indicate the FH prediction of the value of $\chi_{e}N$ at the 
first-order ODT. (The ROL theory cannot be used for this purpose, because it has 
not been used to predict the ODT). The ROL theory predicts a decrease of 10\% - 20\% 
in $q^{\star}$ at the FH ODT, relative to the values at $\chi N = 0$, over a 
surprisingly wide range of molecular weights. This is roughly consistent with the 
range of values found in simulations on values of $\bar{N} \sim 100$ or less and 
in experiment on systems with much higher values of $\bar{N}$: 
$\bar{N} \simeq 10^{3} - 10^{4}$.  
\cite{Bates_Hartney_1985, Bates_Rosedale_1990, Almdal_Bates_1992, 
Rosedale_Bates_1995, Almdal_Bates_1996, Maurer_Bates_Lodge_1998, 
Owens_Russel_1989, Stuhn_Albrecht_1992, Wolff_Ruland_1993, Bartels_Stamm_1994, 
Sakamoto_Hashimoto_1995, Mori_Hashimoto_2001} Quantitative comparisons 
to both simulation and experimental data will be presented elsewhere.

The observation that $q^{\star}$ tends to decrease with increasing $\chi N$
near the ODT has sometimes been described, somewhat loosely, as a result of
``chain stretching'' induced by strong fluctuations near the
ODT.\cite{Almdal_Bates_1990} This description implies that the shift in
$q^{\star}$ is a result of change in intramolecular correlations.  Because the
ROL theory provides predictions of both intramolecular and collective
correlation functions, it is possible for us to ask whether this is the correct
interpretation. Our discussion of this question is based upon the generalized
Ornstein-Zernicke equation, eq. (\ref{OrnsteinZernike}), in which
$cNS^{-1}(q)$ is given as the sum of a term $F(q)$ that depends only upon
single-chain correlation functions and a term $-2  \chi_{a}(q) N$ that is
sensitive to inter-molecular correlations. If the shift of $q^{\star}$ were
primarily a result of changes in the intramolecular correlations, we would
expect it to be well approximated by a modified RPA theory in which Leibler's
approximation for $F(q)$, which assumes Gaussian chain statistics, is replaced
by the function that is obtained from the predicted, slightly non-Gaussian
intra-molecular correlation functions, but in which we still approximate
$\chi_{a}(q)$ by a wavenumber independent value. 

\begin{figure}[tb]
  \centering
  \includegraphics[width=.40\textwidth,height=!]{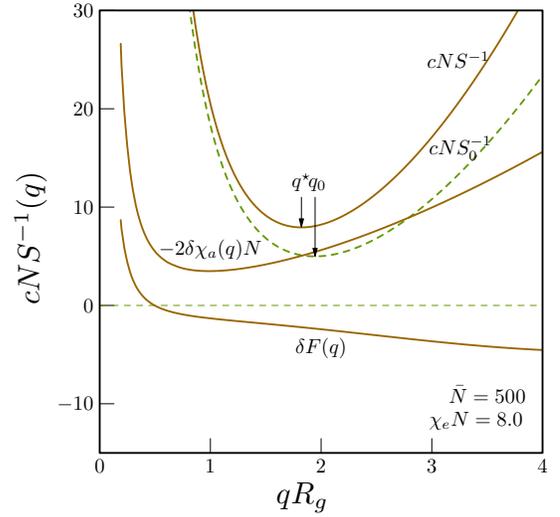}
  \caption[Wavevector dependence of inter- and intra- molecular components to
           the total correlation functions in diblocks.]
  {Wavenumber dependence of inter- and intra- molecular components to the total
  correlation functions from the one loop prediction, evaluated for $\bar N=500$
  and at $\chi_e N = 8.0$. Following eq. (\ref{SinvFchiN}), to display contributions
  from each component, the inverse of $S(q)$ is shown. $c N S_{0}^{-1}$: Leibler's 
  mean field theory prediction. $F(q)$: the intramolecular part with one-loop 
  correction.
  $-2\chi_{a}(q) N$: the intermolecular part with one-loop correction.}
 \label{qdependence}
\end{figure}

To clarify the origin of the dependence of $q^{\star}$ upon $\chi N$,
Fig. \ref{qdependence} shows the $q$-dependence of the predicted inverse 
structure factor $S^{-1}(q)$ and its components, for a system with 
$\bar N = 500$ and $\chi_e N = 8.0$. This figure shows the ROL 
prediction for $S^{-1}(q)$, the RPA prediction $S_{0}^{-1}(q)$, and 
the one-loop contributions $\delta F(q)$ and $-2 \delta \chi_{a}(q)N$ 
to $F(q)$ and $-2\chi_{a}(q) N$, respectively.  For $q$ near $q_{0}$, 
$\delta F(q)$ decreases with increasing $q$, while $-2\chi_{a}(q)N$ 
increases. Ignoring the $q$ dependence of $-\chi_{a}(q)$, as suggested 
above, would thus yield a value of $q^{\star}$ slightly {\it larger} than 
$q_{0}$. This is indeed what is found when $\chi N = 0$, where 
$\delta \chi_{a}(q) = 0$.  The sign of this effect is independent of 
$\chi N$: The contribution to $\delta F(q)$ that arises from deviations 
from Gaussian chain statistics is always a decreasing function of $q$
that would, by itself, always tend to increase $q^{\star}$. The fact that 
$q^{\star}$ drops to values lower than $q_{0}$ near the ODT is a direct
result of the fact that $-2N\chi_{a}(q)$ increases with increasing $q$, 
and that this effect dominates near the ODT. We conclude that the decrease 
in $q^{\star}$ with increasing $\chi N$ is a result of intermolecular,
rather than intramolecular, correlations. Similar conclusions were
reached by Binder and Fried on the basis of Monte Carlo simulations.
\cite{Binder_Fried_1991a}

\begin{figure}[tb]\centering
  \includegraphics[width=.36\textwidth,height=!]{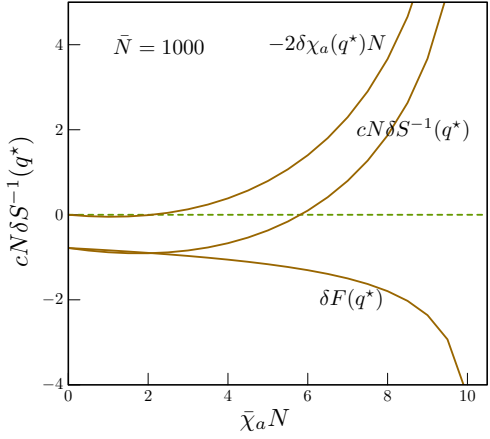}
  \caption[$\barchia N$ dependence of the inter- and intramolecular contributions
   to the inverse of the peak intensity.]
  {The corrections to the inverse of peak intensity evaluated using the ROL
   theory, alongside with the component contributions from inter and intra
   molecular correlations, respectively, versus $\barchia N$, for $\bar N = 1000$.
   The two components have opposite signs near the mean field spinodal.}
  \label{sinvComponents}
\end{figure}

Fig. \ref{sinvComponents} shows values of the ROL predictions to $\delta
F(q^\star)$ and $-2\chi_{a}(q^{\star})N$ at the peak wavenumber $q^{\star}$ as
functions of $\barchia N$, for a melt of symmetric copolymers with
$\bar{N}=1000$. The intramolecular contribution $\delta F(q^\star)$ is
always negative, and thus enhances the peak intensity $S(q^{\star})$, while
$-\delta \chi_{a}(q^{\star}) N$ is positive, and thus suppresses the peak
intensity. The predicted crossover from negative values of $\delta
S^{-1}(q^{\star})$ (enhanced scattering) at low values of $\chi N$ to positive
values (suppressed scattering) closer to the ODT is caused by a change in the
relative importance of these two contributions with increasing $\chi N$. 

%--------------------------------------------------------
\subsection{Peak Intensity (Asymmetric Copolymers)}
\label{subsec:Asymmetric}
%--------------------------------------------------------
The ROL theory may also be used to describe asymmetric copolymers. We 
have shown elsewhere \cite{Morse_Qin_2011} that both the full MFEH 
theory and the ROL theory yield one-loop contributions to $S^{-1}(q)$ 
that, in general exhibit behavior near a spinodal that can be described 
by an expansion of the form
\begin{equation}
    cN \delta S^{-1}(q^{\star}) = 
    \frac{1}{\bar{N}^{1/2}} \left [ 
    \frac{A}{\tau} +
    \frac{B}{\sqrt{\tau}} + C \cdots \right ]
    , \label{dSinvExpansion}
\end{equation}
where the coefficients $A$, $B$ and $C$ depend on copolymer composition 
$f_{A}$ and on the ratio of statistical segment lengths.

The coefficient $A$ of the highest order $1/\tau$ divergence can be 
shown to vanish in the special case of a completely symmetric copolymer 
($f_{A}=f_{B}$ and $b_{A}=b_{B}$) considered by FH, as a result of the
symmetry between the two blocks. These authors took advantage of this 
symmetry by dropping a contribution to the MFEH one-loop theory (a 
Fourier integral) that is zero in this special case, but that actually
yields the dominant $1/\tau$ divergence for asymmetric copolymers. 
It is straightforward to show that the coefficients $A$ and $B$ for
asymmetric copolymers are opposite in sign: The coefficient $B$ that 
is retained in the FH theory is positive, and thus tends to increase 
$S^{-1}(q)$ or decrease $S(q)$, while $A$ is negative for $f_{A} \neq 0$, 
and thus tends to increase $S(q)$ near the spinodal.

\begin{figure}[tb]
\begin{center}
  \subfigure[]{
  \includegraphics[width=0.222\textwidth,height=!]{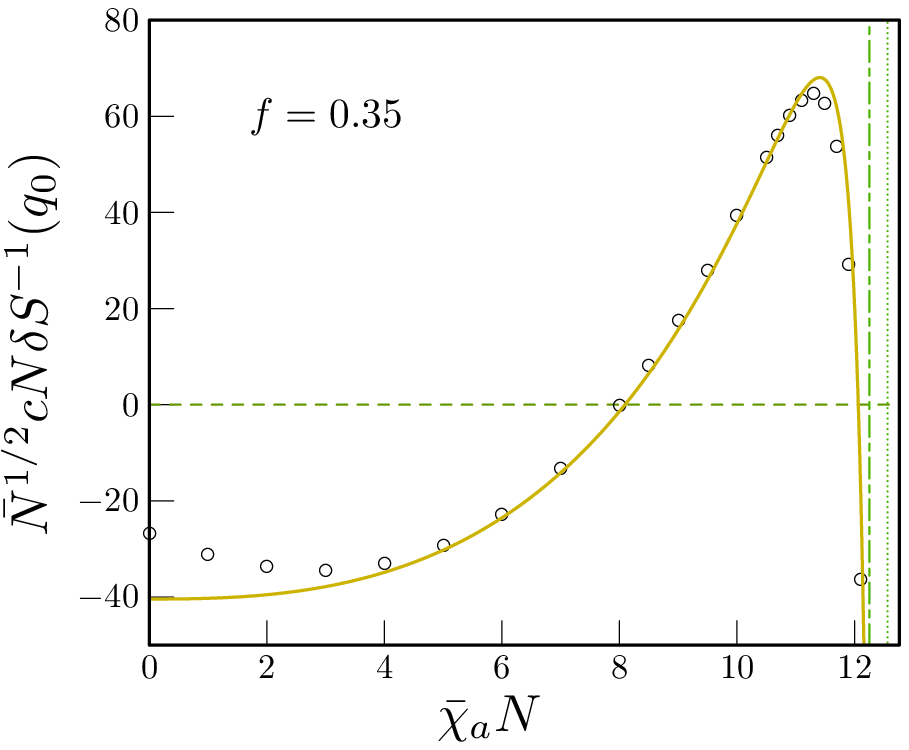}\label{subfig:f035}}
  \subfigure[]{
  \includegraphics[width=0.222\textwidth,height=!]{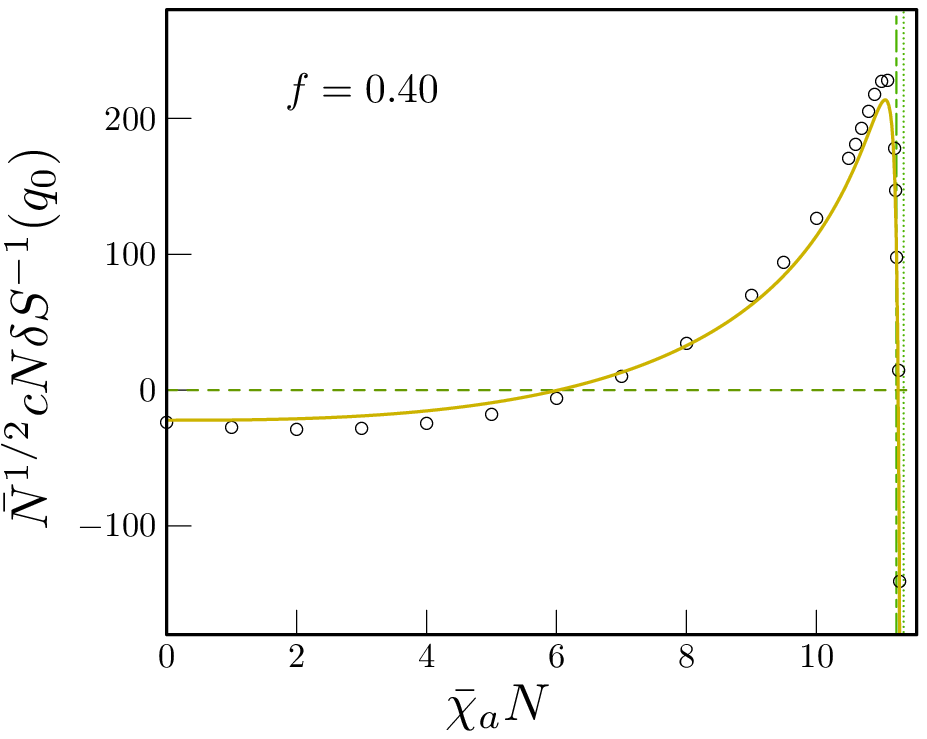}\label{subfig:f040}}
  \subfigure[]{
  \includegraphics[width=0.222\textwidth,height=!]{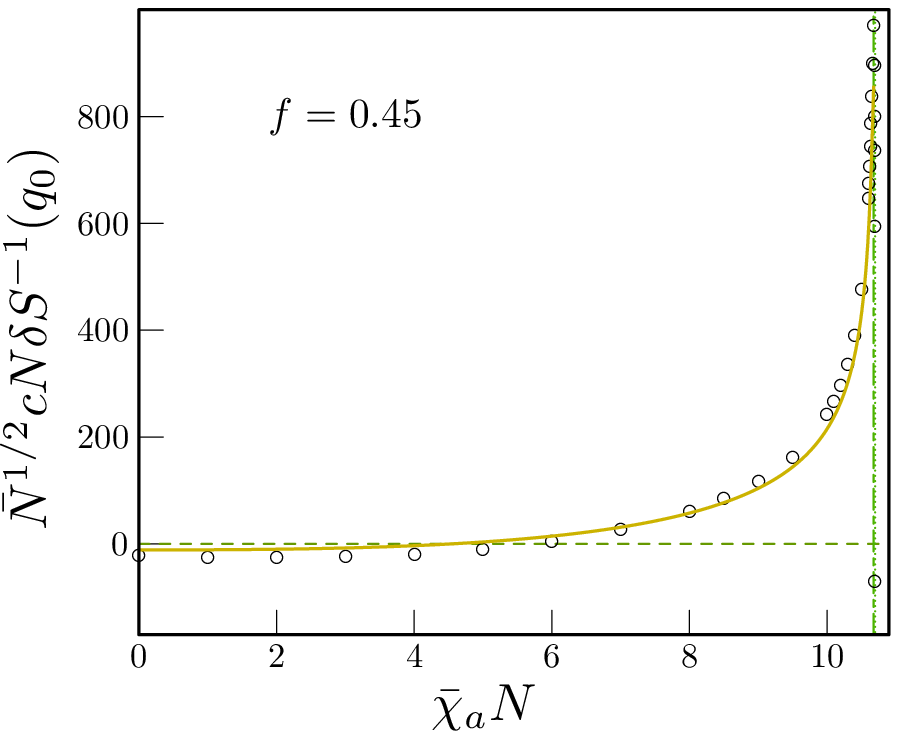}\label{subfig:f045}}
  \subfigure[]{
  \includegraphics[width=0.222\textwidth,height=!]{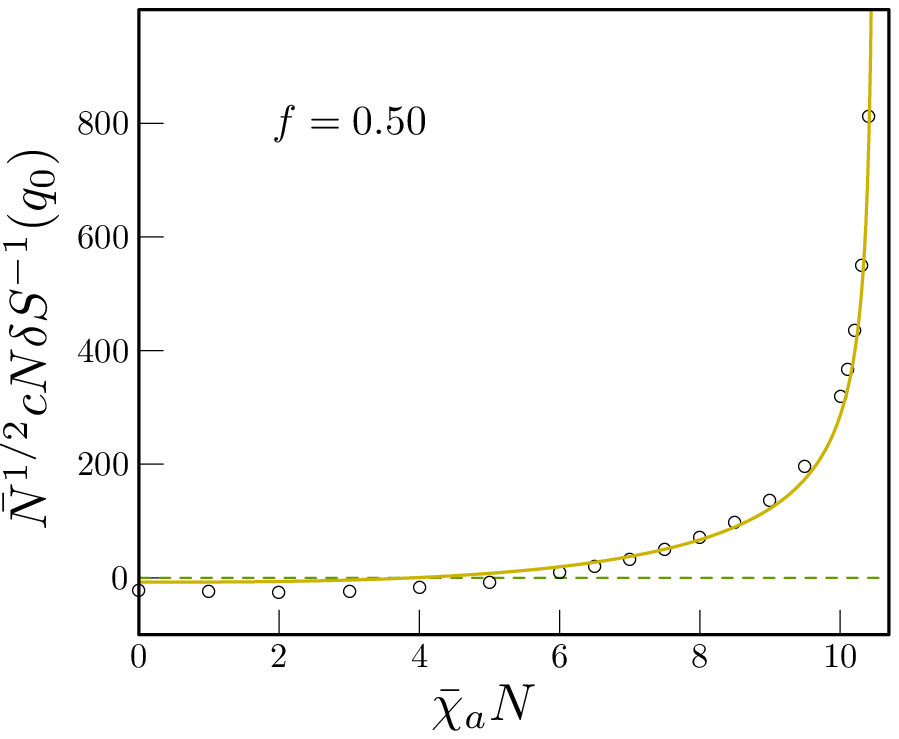}\label{subfig:f050}}
\end{center}
  \caption[$\delta S^{-1}(q_0)$ versus $\barchia$ for asymmetric
diblocks.]{$\delta S^{-1}(q_0)$ versus $\barchia$ for diblock copolymer with
  $b_A = b_B$, at various block fractions: $f = 0.35$, 0.40, 0.45 and 0.50.
  Solid lines are fitted results using eq. (\ref{dSfit}). The dash-dotted and
  dotted perpendicular lines are ODTs calculated from SCFT and the mean field
  spinodal, respectively.}
  \label{fig:dSfit}
\end{figure}

Fig. (\ref{fig:dSfit}) shows results of the ROL theory for the 
normalized one-loop contribution $\bar{N}^{1/2}cN \delta S^{-1}(q_{0})$
for modestly asymmetric and symmetric diblock copolymers with 
$b_{A} = b_{B}$, for systems with $f_{A} =$ 0.35, 0.40, 0.45, and 0.5. 
In the case of a symmetric polymer, the quantity 
$\bar{N}^{1/2}cN \delta S^{-1}(q_{0})$ increases monotonically with 
increasing $\barchia N$, as also shown Fig. \ref{asymptotics}. For the
most asymmetric case of $f_{A} = 0.35$, the one-loop contribution is 
clearly a non-monotonic function of $\barchia N$, which increases with 
increasing $\barchia N$ far from the spinodal, but then decreases 
rapidly near the spinodal.  This is a result of the competition 
between the $A/\tau$ term with $A < 0$, and the $B/\sqrt{\tau}$ term, 
with $B > 0$, in eq. (\ref{dSinvExpansion}). 

For asymmetric copolymers, SCFT predicts a first-order ODT. The two vertical 
dotted lines in each figure are SCFT predictions for the values of $\chi N$
at the spinodal (the larger value) and at the ODT (the smaller value). For 
$f_{A} = 0.5$, these values are equal. Note that, for the most asymmetric 
copolymer, with $f_{A}=0.35$, the maximum in $\bar{N}^{1/2}cN 
\delta S^{-1}(q_{0})$ appears at a value of $\chi N$ significantly below 
the SCFT ODT value. The theory thus suggests that strong fluctuation effects 
that are qualitatively different from those found for symmetric copolymers 
could appear within the disordered phase of asymmetric copolymers. 

\begin{figure}[tb]
  \centering
  \includegraphics[width=.37\textwidth,height=!]{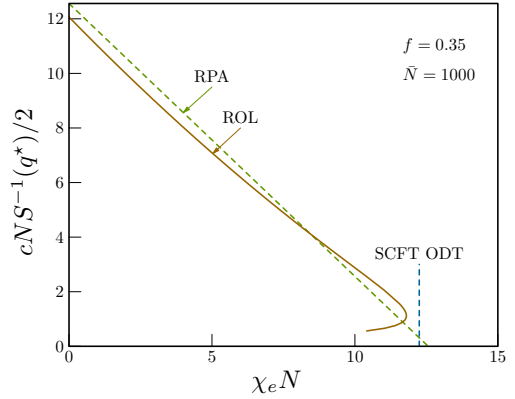}
  \caption{Self-consistently calculated inverse peak intensity versus $\chi_e N$
   for diblock copolymer with $f=0.35$ and $\bar N = 1000$. Predictions
   of RPA, and the ROL theories are shown as dashed and solid lines, respectively.
   ODT calculated from the SCFT is also marked.}
  \label{fig:sinvf35}
\end{figure}

In the results shown in Fig. (\ref{fig:dSfit}), the input parameter
$\chi N$ may be calculated using either the SCFT interaction parameter
$\chi_{e}$, to obtain a perturbative form of the ROL theory, or using the 
apparent interaction parameter $\bar{\chi}_{a}$, to obtain a 
self-consistent Hartree approximation. The distinction between these
two approaches is unimportant far from the ODT, where both yield small
corrections to the RPA, but become important near the ODT, where the
predicted corrections become large. Brazovski\u{\i} argued that the
Hartree approximation should be accurate near the crystallization 
transition in a class of systems with a scalar order parameter field 
that exhibits strong fluctuations at a nonzero wavenumber $q^{*}$,
and in which a Taylor expansion of the effective Hamiltonian has a 
vanishing cubic term (leading to a second-order transition in Landau 
theory) and a small quartic term.  His analysis justifies the use of 
a Hartree approximation for symmetric copolymers, for which the cubic 
term in the Landau theory must vanish by symmetry, in the limit
$\bar{N} \rightarrow \infty$, in which the effects of the quartic
anharmonicity become small. It does not, however, justify the use 
of a Hartree approximation for strongly asymmetric copolymers. 

Fig. \ref{fig:sinvf35} shows the results obtained if (despite the
above discussion) we use the self-consistent (Hartree) ROL theory
to calculate $S(q^{*})$ for asymmetric diblocks with $f=0.35$ and 
$\bar{N}=1000$. For sufficiently small values of $\chi_{e} N$, 
the behavior is similar to that shown in Fig.  \ref{scfsinv} for 
symmetric diblocks.  For $\chi_e N$ close to 12, however, we see 
an unphysical turning point, leading to a predicted value of 
$cN S^{-1}(q^{*})$ that is a multivalued function of $\chi_e N$. 
The behavior is similar to that obtained in previous attempts to
use a Hartree approximation to describe the peak scattering 
intensity at $q = 0$ in off-critical polymer blends.
\cite{Wang_2002, Qin_Morse_2009} In the example shown here, the 
turning point occurs at a value of $\chi_{e} N$ that is below 
both the SCFT prediction of the ODT for this value of $f_{A}$ 
and below the FH prediction of $(\chi_{e} N)_{c} \simeq 14.6$ 
for the ODT of a symmetric copolymers with $\bar{N} = 1000$.
Both of these values are presumably less than the true value 
of $\chi_{e}N$ at the ODT for this system. This indicates that 
the self-consistent ROL is almost certainly not an adequate 
description of moderately asymmetric copolymers near the ODT. 
We do, however, expect either form of ROL to accurately 
describe small corrections to the RPA farther from the ODT.

%---------------------------------------
\subsection{Molecule and Block Dimensions}
\label{subsec:size}
%---------------------------------------

Previous theoretical and simulation studies have characterized the
statistics of individual chains primarily by examining how the radii 
of gyration of entire chains and of the A and B blocks change with 
$\chi N$. Let $R_{g}$ be the radius of gyration of an entire copolymer, 
and let $R_{g,i}$ be the radius of gyration of the $i$ block ($i=$A or 
B).  Let $R_{AB}^{2}$ be the mean-squared distance between the centers 
of mass of the A and B blocks. These quantities are related to each 
other by
\begin{equation}
   R_g^2 = 
   f_A R_{g,A}^2 + f_B R_{g,B}^2 + f_A f_B R_{AB}^2 .
   \label{sizeIdentity}
\end{equation}
They, respectively, are also related to the low-$q$ behavior of the
intramolecular correlation functions $\Omega_{ij}(q)$, in the Guinier 
regime $q R \ll 1$, by relations of the form
\begin{equation}
   \lim_{q R \ll 1}
   \Omega_{ij}(q) \simeq  c N f_{i}f_{j}
   \left [ 1 - \frac{1}{3} q^{2} L_{ij}^{2} \right ]  ,
\end{equation}
where we have
\begin{align*}
   L_{AA}^{2} & =  R^{2}_{g,A} , \\
   L_{AB}^{2} & =  \frac{1}{2}
   \left( R_{g,A}^{2} + R_{g,B}^{2} + R_{AB}^{2} \right) ,
  \nonumber 
\end{align*}
and an analogous expression for $L_{BB}^{2}$.
Using these expressions, it is straightforward to show that $F(q)$
diverges like $1/q^{2}$ as $q \rightarrow 0$ and that
\begin{equation}
   \lim_{q R \ll 1}
   F^{-1}(q) \simeq \frac{1}{3} f_{A}^{2} f_{B}^{2} R_{AB}^{2} q^{2}, 
   \label{FinvGuinier}
\end{equation}
in the same regime.
%\begin{eqnarray}
%  |\Omega| \simeq \left(\frac{N f_A f_B}{ v}\right)^2 \frac{q^2}{3} R_{AB}^2 .
%  \label{eq:dbcRAB}
%\end{eqnarray}
%This also implies, in the low $q$ regime, $S(q, \chi_e = 0) = |\Omega| / \Omega_+ \simeq N
%f_A^2 f_B^2 q^2 R_{AB}^2 / (3v)$.
%For non-zero $\chi_e$, the $1/q^2$ term in $\chi_{a} N$ gives additional contribution
%to low-$q$ expansion of $S$. For ideal chains, $R_{AB}^2 = 2 (R_{g,A}^2 +
%R_{g,B}^2)$. For ideal symmetric chains, $R_{g,A}^2 = R_{g,B}^2 = R_g^2 /2$ and
%$R_{AB}^2 = 2 R_g^2$.

\begin{figure}[tb]
  \centering
  \includegraphics[width=.40\textwidth,height=!]{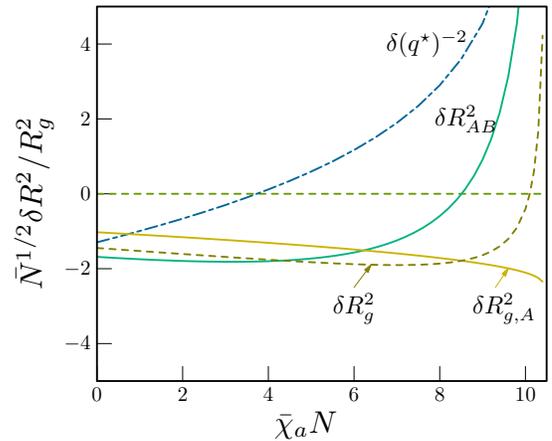}
  \caption[Size of molecule and blocks for diblocks of varying $f_A$.]
  {$\chi$ dependence of the fractional change in the radius of gyration of one 
  block ($R_{g,A}^2$) and of the whole molecule ($R_g^2$), and the center of mass 
  distance of two blocks ($R_{AB}^2$), for a symmetric diblock. The curve 
  labelled with $\delta(q^\star)^{-2}$ represents the corresponding 
  fractional change $\bar{N}^{1/2}((q_0/q^\star)^2 - 1)$ in the length 
  scale $(q^\star)^{-1}$. }
  \label{fig:sizeSym}
\end{figure}

\begin{figure}[tb]
  \subfigure[]{
  \includegraphics[width=0.222\textwidth,height=!]{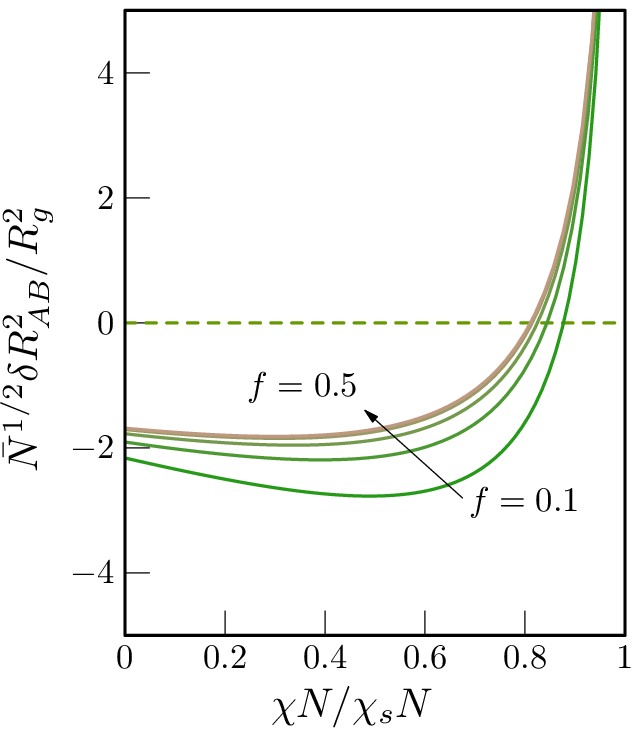}
  \label{subfig:dRAB}}
  \subfigure[]{
  \includegraphics[width=0.232\textwidth,height=!]{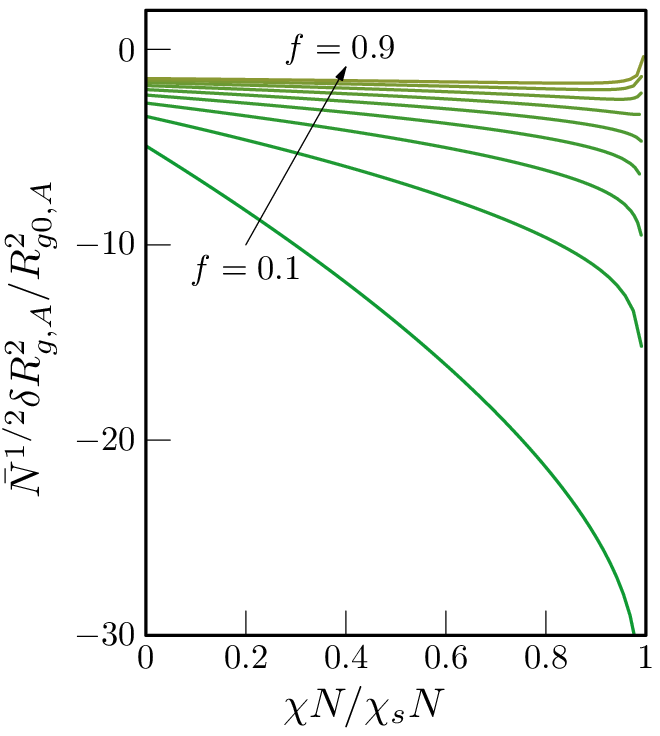}
  \label{subfig:dRgA}}
  \caption[Composition and $\chi$ dependence of diblock dimensions.]
  {The change of mean squared distance of center of mass
  of two blocks ($R_{AB}^2$, left) and the radius of gyration of one block
  ($R_{g,A}^2$, right), for block fractions $f = 0.1$, 0.2, ..., 0.9. The
  abscissa is $\chi N$ normalized by the mean field
  spinodal.}\label{fig:sizeAsym}
\end{figure}

Fig. \ref{fig:sizeSym} shows the $\chi$ dependence of the one-loop 
corrections to $R^{2}_{g}$, $R_{g,A}^{2}$ and $R_{AB}^{2}$ for a symmetric 
diblock copolymer.  Because the predicted fractional changes are all 
proportional to $\bar{N}^{-1/2}$, we show fractional changes multiplied by 
$\bar{N}^{1/2}$ to obtain quantities that are independent of chain length.
 
At $\chi = 0$, the calculated corrections $R_{g,i}^{2}$, $R_{AB}^{2}$ 
and $R_{g}^{2}$ are all negative. The predicted change in the overall 
radius of gyration, $\delta R_g^2 / R_g^2 = - 1.42 \bar N^{-1/2}$ at 
$\chi N = 0$, is the same as that found previously for monodisperse 
homopolymers.\cite{Qin_Morse_2009} 
%The negative correction to $\delta R_{AB}^{2}$ at
%$\chi N = 0$ causes a negative correction to $S(q)$ in the low-$q$ 
%regime, shown as a positive correction in Fig. \ref{deltaF} for 
%$\delta F(q)$, since $S(q) \propto R_{AB}^{2}q^{2}$ at low $q$ when 
%$\chi_{a}(q) = 0$.  
As already noted, these predicted deviations in chain dimensions are
all defined as deviations from the predictions of a random walk model 
in which the statistical segment length is defined to be that of a 
hypothetical system of infinite chains. Because the deviation is 
negative and decreases as $\bar{N}^{-1/2}$ with increasing $N$, it 
implies that longer chains are slightly more swollen than shorter 
ones. The reasons for this $N$-dependent deviation from random walk 
statistics has been discussed previously by Beckrich {\it et al.}
\cite{Beckrich_Wittmer_2007} 

As $\chi$ increases, the mean-squared distance $R_{AB}^{2}$ between 
the block centers of mass increases, while the block radii of 
gyration decrease slightly. This qualitative trend has been 
previously predicted \cite{Barrat_Fredrickson_1991} and observed 
in simulations.\cite{Binder_Fried_1991a} The phenomena is consistent 
with the idea that the two blocks avoid increasingly unfavorable AB 
contacts by shrinking and by moving away from one another. The deviation
of the total radius of gyration from the random walk value is negative 
and almost independent of $\barchia N$ for $\barchia N \lesssim 9$, but 
increases rapidly with $\barchia N$ near the spinodal.  The predicted 
corrections to all these quantities diverge like $\tau^{-1/2}$ close 
the spinodal, as does the correction to the inverse peak intensity. 
These divergences are pre-empted by the appearance of a first-order 
transition.  For $\bar N = 500$, the predicted fractional changes in 
$R_g$, $R_{AB}$, $R_{g,A}$ between $\chi N = 0$ and $\chi N = 10$ 
are 2.5\%, 9.4\% and -5.5\%, respectively.

For comparison, Fig. \ref{fig:sizeSym} also shows the fractional change 
in the squared peak wavelength $1/(q^{\star})^{2}$ as a function of 
$\barchia N$, which increases monotonically with increasing $\chi N$.  
Note that the fractional change in $1/q^{2}$ is predicted to increase 
notably more rapidly with increasing $\barchia N$ than either $R_{AB}^{2}$ 
or $R^{2}_{g}$, in agreement with simulation results. This comparison 
is enough to show that it is impossible to quantitatively account for 
the shift in value of the nonzero peak wavenumber $q^{\star}$ by 
examining overall chain dimensions or (equivalently) the low-$q$ 
behavior of the single-chain correlation function. The above analysis 
of the behavior of $F(q)$ for $q$ near $q^{\star}$ indicates, moreover, 
that any attempt to interpret the shift in $q^{\star}$ as a direct 
result of change in intramolecular correlations is actually qualitatively 
wrong.

Fig. \ref{fig:sizeAsym} shows how the same measures of chain and block
sizes depend on copolymer composition $f_{A}$ in melts of asymmetric 
copolymers. Fig. \ref{subfig:dRAB} shows the results for $\delta R_{AB}^2$ 
and Fig. \ref{subfig:dRgA} for $\delta R_{g,A}^2$.  In both plots, the $x$ 
axis is the ratio of $\barchia N$ to its spinodal value. $\delta R_{AB}^2$ 
shows a relatively weak dependence on $f$.  For highly asymmetric
copolymers, the radius of gyration of the minority block shrinks much 
more than the majority block, and the majority block increases with 
increasing $\barchia N$ near the spinodal.

%---------------------
\section{Conclusions}
\label{sec:conclusion}
%---------------------

We have presented quantitative predictions of the ROL theory for corrections 
to the RPA description of correlations in disordered diblock copolymer melts.  
The theory is expected to have a wider range of validity than the earlier FH 
theory and its descendants, and makes independent predictions about 
intramolecular and collective correlation functions. 

When applied to symmetric diblock copolymers, the ROL theory yields predictions
that are similar to those of the FH and BF theories near the ODT, but that are
qualitatively different in the limit of low $\chi N$ values. The ROL theory
predicts a depression of the peak scattering intensity $S(q^\star)$ near the
ODT but a slight enhancement in $S(q^\star)$ for $\chi N \alt 6$, and decrease
in $q^{\star}$ near the ODT but a slight increase at small values of $\chi N$,
relative to RPA predictions. In the limit $\chi N = 0$, these deviations from
RPA predictions are direct results of the deviations from Gaussian
single-chain statistics in a dense polymer liquid that have been studied
previously by the Strasbourg
group.\cite{Beckrich_Wittmer_2007,Wittmer_Beckrich_2007} Near the ODT,
corrections to the RPA are instead dominated by corrections to the apparent
interaction parameter $\chi_{a}(q)$.

Like the BF theory, the ROL theory predicts a monotonic decrease in 
$q^{\star}$ with increasing $\chi N$. The ROL theory also predicts a 
shrinkage of individual blocks, and an increase in the distance between 
the A and B blocks, in agreement with simulation results. We have 
tried to further clarify the physical origin of the shift in $q^{\star}$ 
by comparing the results of the full theory to those of a simplified 
theory in which we take into account the predicted deviations from 
single-chain statistics, via the function $F(q)$, but approximate the 
apparent interaction parameter $\chi_{a}(q)$ by a parameter that is 
independent of $q$. In this simplified theory, the effect of deviations 
from Gaussian statistics alone would actually be to {\it increase} 
$q^{\star}$ above the value predicted by the RPA, as is found in the regime
$\chi N \lesssim 6$ in which this intramolecular effect dominates the 
correction to $S(q)$. The decrease in $q^{\star}$ with increasing $\chi N$ 
is instead found to be a result of the $q$-dependence of the apparent 
interaction parameter, and is thus best understood as a result of 
changes in intermolecular rather than intramolecular correlations.

When applied to asymmetric diblock copolymers, the ROL theory predicts a 
correction to the RPA prediction for $S^{-1}(q^{\star})$ that increases with 
increasing $\chi N$ far from the spinodal transition, but that reaches a 
maximum and then begins to decrease very near the spinodal. Applying the 
``Hartree" approximation that underlies the Brazovski\u\i and FH theories to
asymmetric polymers yields strikingly unphysical results near the ODT. 

A comparison of the predictions of the ROL and FH theories for symmetric
copolymers over the range of parameters examined in the neutron scattering 
data of Bates, Rosedale and Fredrickson \cite{Bates_Rosedale_1990} suggests 
that it would be difficult to discriminate between these two theories on the 
basis of these experiments, which were carried out over a limited range of 
values of $\chi N$ in a model system of unusually high molecular weight. We 
will compare both theories to the results of computer simulations of shorter
polymers over a wide range of values of $\chi N$ in a subsequent paper. 

% ------------------------------------------------------------------------------
\appendix
% ------------------------------------------------------------------------------

% ---------------------------------
\section{Renormalization Procedure}
\label{appdx:renormalization}
% ---------------------------------

The results presented in this paper are all obtained by applying the renormalization 
scheme that was outlined in ref. \cite{Piotr_Morse_2007} to the Fourier integrals 
that appear in eqs. (\ref{intradef}) and (\ref{sigmadef}). In this scheme, the UV 
divergent parts of the Fourier integrals that contribute to $S^{-1}(q)$ are all 
identified with renormalization of the interaction and statistical segment length 
parameters, and thus implicitly absorbed into changes in the values of these
parameters, while the UV convergent parts are identified as corrections to the 
RPA. To extract the UV convergent part of such integrals, we have numerically 
evaluated each integral with respect to a wavevector $\kv$ over a domain 
$|\kv| < \Lambda$, for $\Lambda R \gg 1$, and then subtracted off the 
analytic result obtained in ref. \cite{Piotr_Morse_2007} for the UV divergent 
part of the corresponding integral. In each case, the resulting difference is found to
be approximately independent of $\Lambda$ for $\Lambda R \gg 1$, and to approach 
a finite limit as $\Lambda R \gg \infty$, confirming our analytic calculations
of the UV divergent contributions. Our final estimate for the UV convergent
part of each such integral is obtained by repeating this procedure for several 
values of $\Lambda$ and then numerically extrapolating to $\Lambda = \infty$. 
The integrals with respect to $\kv$ in eqs. (\ref{intradef}) and (\ref{sigmadef})
were calculated with a 2D Rhomberg integration algorithm,\cite{NR_Fortran_1992} 
using the wavenumber $k = |\kv|$ and the cosine of the angle between $\kv$ and 
$\qv$ as coordinates. 

A similar procedure was used in ref. \cite{Qin_Morse_2009} to calculate the
one-loop corrections to the free energy density and to the $\qv \rightarrow 0$
limit of $\chi_{a}(\qv)$ in a polymer blend. These calculations only required
one-dimensional integrals with respect to $|\kv|$. 

\emph{Intramolecular Correlations} 
It was shown in ref. \cite{Piotr_Morse_2007} that the result of eq. (\ref{intradef}) 
for $\Ss_{ij}(\qv)$ can be expressed as a sum
\begin{equation}
  \Ss_{ij}(\qv) = \Ssi_{ij}(\qv) 
  + \Ssi_{ij}^{(\Lambda)} + \delta\Ss_{ij}(\qv) ,
\end{equation}
in which $\Ssi_{ij}^{(\Lambda)}$ is a UV divergent contribution that was found to 
be of the form
\begin{equation}
  \Ssi_{ij}^{(\Lambda)} = 
  \frac {\partial \Ssi_{ij}(\qv;b)} {\partial b_k} \delta b_k (\Lambda)
  , \label{SsiLambda}
\end{equation}
where 
\begin{equation}
   \delta b_k (\Lambda) = l_k^2 \Lambda / (\pi^2 \bar{l})
   \label{deltaB}
\end{equation} 
is a UV divergent shift in the value of the statistical segment length $b_{k}$.
Here, $\bar{l} = f_A l_A + f_B l_B$, where $l_{i} = v/b_{i}^{2}$. The function
$\Ssi_{ij}(\qv;b)$ denotes the random-walk model for $\Ssi_{ij}(\qv)$ for a 
chain with specified statistical segment lengths $b_{A}$ and $b_{B}$. The UV
divergent term was thus interpreted as the first term in an expansion of the
intramolecular correlation function $\Ssi_{ij}(\qv,b+\delta b)$ for a random 
walk with renormalized statistical segment lengths. The UV convergent 
term $\delta\Ss_{ij}(\qv)$ is calculated by subtracting the result
of eqs. (\ref{SsiLambda})-(\ref{deltaB}) for the UV-divergent part of the 
integral from a numerical result for the full integral. 

\emph{Direct Correlations}
The one-loop correction to $\chi_{a}(\qv)$ is obtained from eqs. (\ref{interdef})
and (\ref{sigmadef}). It was shown in ref. \cite{Piotr_Morse_2007} that this 
quantity can be expressed as a sum of the form
\begin{equation}
  \chi_{a}(\qv) = \chi_{e}(\Lambda) + \frac{H(\qv R)}{N}\Lambda + \delta\chi(\qv)
  ,
\end{equation}
with a cutoff-dependent effective interaction parameter $\chi_{e}(\Lambda)$
of the form
\begin{equation}
  \chi_{e}(\Lambda) = \chi_0 
  + A\Lambda^3 + B\chi_0 \Lambda 
  . \label{chiedef}
\end{equation}
The coefficients $A$ and $B$ are independent of $\Lambda$, $N$ and $q$, 
but depend upon the parameters $b_{A}$, $b_{B}$, $v$ and $f_{A}$. The 
coefficient $H(\qv R)$ depends upon the normalized wavenumber $\qv R$ as well
as $b_{A}$, $b_{B}$, $v$ and $f_{A}$.  Explicit expressions for $A$, $B$, 
and $H$ are given in ref.  \cite{Piotr_Morse_2007}.  We argued there
that the $\Lambda$-dependent terms in eq. (\ref{chiedef}) were naturally 
interpreted as trivial corrections to the value of the local interaction 
parameter, because they are found to be independent of $N$ and $q$, but
do depend on parameters that reflect the local liquid correlations of this
simple model. By the same reasoning, the contribution $H(\qv R)\Lambda/N$
cannot be interpreted as a simple renormalization of $\chi$, because it 
depends on both $\qv$ and $N$. It was shown, however, that this term could 
be attributed to corrections to the RPA that arise from: 
i) perturbations in local liquid structure near chain ends and near the 
junction between the A and B blocks, and ii) a gradient-squared contribution 
to the excess free energy functional. It was also found, however, that 
the coefficients $H(\qv R)$ and $A$ both vanish in the one-loop approximation
in the special case $b_{A}=b_{B}$.  The results presented in this paper 
are all obtained for systems with $b_{A} = b_{B}$, for which the only UV divergent 
contribution to $\chi_{a}(\qv)$ is a constant $B\chi_{0}\Lambda$.  
The UV-convergent contribution $\delta\chi_{a}(\qv)$ has thus been 
calculated by subtracting this contribution from a numerical result for 
the full one-loop correction $\Delta\chi_{a}(\qv, \Lambda)$.

% ----------------------------------
\section{Ideal Symmetric Copolymers}
\label{app:zeroChi}
% ----------------------------------

Here, we show that $\chi_{a}(\qv) = 0$ for a completely symmetric diblock 
copolymer with $\chi_{0}=0$.  We first give a general symmetry argument as 
to why this must be true in this special case, and then confirm explicitly 
that this is consistent with results obtained from the one-loop theory.

{\it General Symmetry Argument}:
Consider the relationship between the correlation function matrix
$\Sc_{ij}(\qv)$ in a liquid of diblock copolymers containing physically 
identical $A$ and $B$ blocks of equal lengths and the correlation function 
$S(\qv)$ in a corresponding homopolymer liquid, in which we make no
distinction between the two blocks. In the homopolymer melt, 
\begin{equation}
   S(\qv) = \Ss(\qv) + H(\qv) ,
   \label{SIntraInterHomo}
\end{equation}
where $\Ss(\qv)$ and $H(\qv)$ are the intra- and inter-molecular 
pair correlation functions, respectively, in the liquid with no
differential labelling. If each molecule in this liquid is instead
treated as a diblock, we obtain a matrix of correlation functions
\begin{equation}
    \Sc_{ij}(\qv) = \Ss_{ij}(\qv) + \frac{1}{4} H(\qv) ,
   \label{SIntraInterDiblock}
\end{equation}
where $\Ss_{ij}(\qv)$ is a matrix of intramolecular correlation 
functions, and $H(\qv)/4$ is an intermolecular contribution, in 
which $H(\qv)$ is the function defined for a one-component liquid
in eq. (\ref{SIntraInterHomo}). The scalar intramolecular 
correlation function $\Ss(q)$ of eq. (\ref{SIntraInterHomo})
is given by the sum $\Ss(\qv) = \sum_{ij}\Ss_{ij}(\qv)$. The 
intermolecular contribution to $\Sc_{ij}(\qv)$ is simply $H(\qv)/4$ 
in eq. (\ref{SIntraInterDiblock}) because each inter-molecular 
pair in the one-component liquid has an equal probability of being 
labelled AA, BB, AB or BA by a process in which one end
one end of each chain is labelled A and the other B at random. 

Because the matrices $\Sc_{ij}(\qv)$ and $\Ss_{ij}(\rv)$ must 
be invariant under an exchange of the labels A and B,
the eigenvectors of $\Sc_{ij}(\qv)$ and $\Ss_{ij}(\qv)$ must thus 
be proportional to the even and odd vectors 
\begin{eqnarray}
    \delta   & \equiv & (1,1) , \nonumber \\
    \varepsilon & \equiv & (1,-1) .
\end{eqnarray}
It follows that $\Sc_{ij}(\qv)$ may be expressed as a sum
\begin{eqnarray}
   \Sc_{ij}(\qv) & = &
   \Sc_{+}(\qv)\delta_{i}\delta_{j}/2 +
   \Sc_{-}(\qv)\varepsilon_{i}\varepsilon_{j}/2 ,
\end{eqnarray}
in which $\Sc_{+}(\qv)$ and $\Sc_{-}(\qv)$ are its eigenvalues. 
Let $\Ss_{+}(\qv)$ and $\Ss_{-}(\qv)$ be the corresponding
eigenvalues of $\Ss_{ij}(\qv)$.  By contracting eq. (\ref{SijDef}) 
for $\Sc_{ij}(\qv)$ with $\varepsilon$ to project onto the odd 
subspace, it is straightforward to show that
\begin{equation}
   \Sc_{-}(\qv) = 
  \frac{1}{2}\sum_{ij}\Sc_{ij}(\qv)\varepsilon_{i}\varepsilon_{j}
  = \Ss_{-}(\qv)
  .
\end{equation}
By projecting the Ornstein-Zernicke equation for $S^{-1}_{ij}(\qv)$
onto the odd subspace, and using the fact that
$\Sc_{-}^{-1}(\qv) = \Ss_{-}^{-1}(\qv)$, we found
\begin{equation}
   \sum_{ij}C_{ij}(\qv)\varepsilon_{i}\varepsilon_{j} = 0 .
\end{equation}
It follows immediately from eq. (\ref{chiadef}) that 
$\chi_{a}(\qv)=0$ in the incompressible limit.

{\it One-Loop Theory}:
To calculate $\delta\chi_{a}(\qv)$ in the one-loop theory, we start 
from eqs. (\ref{interdef}) and (\ref{sigmadef}) for $\delta C_{ij}(\qv)$. 
When $\chi_{0}=0$, eq. (\ref{Gdef}) for $\Gh_{ij}(\qv)$ 
reduces to a scalar $\Gh(\qv) = 1/\Sssum(\qv)$. As a result, the 
summation over $k,m$ and $l,n$ in eq. (\ref{sigmadef}) can be carried 
out separately. By defining 
\begin{equation}
   \bar{\Ss}_i (\qv, \kv) 
   \equiv \sum_{klm} \Ssi^{-1}_{ik}(\qv)
   \Ssi^{(3)}_{klm}(\qv, \kv_-, -\kv_+)
   ,
\end{equation}
we may rewrite eq. (\ref{interdef}) as
\begin{equation}
    \delta C_{ij}(\qv) = \frac{1}{2} \int_\kv
    \bar{\Ss}_i(\qv, \kv) \bar{\Ss}_j(\qv, \kv) \Gh^{-1}(\kv_+)
    \Gh^{-1}(\kv_-)
    , \label{deltaC_chiZero}
\end{equation}
where we have used the fact that 
$\bar{\Ss}_{i}(\qv, \kv) = \bar{\Ss}_{i}(-\qv, -\kv)$.

For a symmetric diblock copolymer, however, we also have
$\bar{\Ss}_A(\qv, \kv) = \bar{\Ss}_B(\qv, \kv)$,
as a result of the symmetry under an exchange of the labels of 
the A and B monomers. Using this in eq. (\ref{deltaC_chiZero})
yields an expression for $\delta C_{ij}(\qv)$ that is independent
of the indices $i$ and $j$. 

\section{Empirical Approximations}
\label{app:asymptoticS}
To facilitate comparison with experimental and simulation results, 
we have developed empirical approximations for some of our results
for polymers with $b_{A} = b_{B}$. These are presented in this
appendix.

\subsection{Peak Intensity}
The behavior of $c N \delta S^{-1}(q_{0})$ for diblock copolymers 
with $b_{A} = b_{B}$ but modestly asymmetric compositions is 
reasonably well approximated for $0.35 < f < 0.65$ by a function
\begin{align}
         & cN \delta S^{-1} (q_0) \\
  \simeq & \frac {1} {\bar N^{1/2}} \left[ 
     \frac {a(f) (\chi N)^c } {\sqrt{\chi_s N - \chi N}} 
   + \frac {b(f) (\chi N)^c } { \chi_s N - \chi N}
   + d(f) \right] \nonumber
  \label{dSfit}
\end{align}
with composition dependent coefficients
\begin{align}
  a(f) &=  0.652 - 0.799 \left( \frac{1}{2} - f \right)^2 , \nonumber\\
  b(f) &=        - 18.2  \left( \frac{1}{2} - f \right)^2 , \nonumber\\
  c(f) &=  2.50  + 0.857 \left( \frac{1}{2} - f \right)^2 , \nonumber\\
  d(f) &= -7.58  - 1460  \left( \frac{1}{2} - f \right)^2 .
\end{align}
The expression is symmetric with respect to $f \rightarrow 1-f$, and
the coefficient $b(f)$ of the strongest divergence vanishes at $f=1/2$.
The quality of the fits is illustrated in Fig. \ref{fig:dSfit}.

%An analogous fit for symmetric binary homopolymer blends contains both the low
%$\chi$ and large $\chi$ asymptotics\cite{Qin_Morse_2009}: $\bar{N}^{1/2} \delta
%\chi N \simeq h_0(x) = (6/\pi)^{3/2} \chi N$ for small $\chi$ and $\simeq
%h_1(x) = -3.71 + 27(2-\chi N)^{1/2}/\pi$ for $\chi N$ close to 2. By
%introducing the appropriate weight, the curve in Fig.2 of \cite{Qin_Morse_2009}
%can be fitted by:
%\begin{equation}
%  h_0(x) \left( 1 - \chi N/2 \right)^{m_1} +
%  \left(h_1(x) - m_2(2-\chi N)\right) \left( \chi N/2 \right)^{m_3}
%\end{equation}
%with $m_1 = 1.30$, $m_2 = 3.86$, $m_3 = 1.30$.

% -------------------------------------------------------
\subsection{Peak Wavenumber}
\label{app:asymptoticQ}
% -------------------------------------------------------

Let $f(x,y, \bar{N})$ denote the quantity $c N S^{-1}(q)$, where 
$x \equiv qR_{g,0}$ and $y \equiv \chi N$.  The ROL theory prediction 
for this quantity is a sum of the form
\begin{equation}
  f(x, y, \bar{N}) = f_0(x,y) + \frac{f_1(x,y)}{\bar{N}^{1/2}},
\end{equation}
in which $f_{0}(x,y) = cNS_{0}^{-1}(qR, \chi N)$ is the RPA 
prediction, and $f_{1}(x,y)/\bar{N}^{1/2}$ is the one loop 
contribution.

\begin{figure}[tbh]
\centering
\includegraphics[width=0.3\textwidth,height=!]{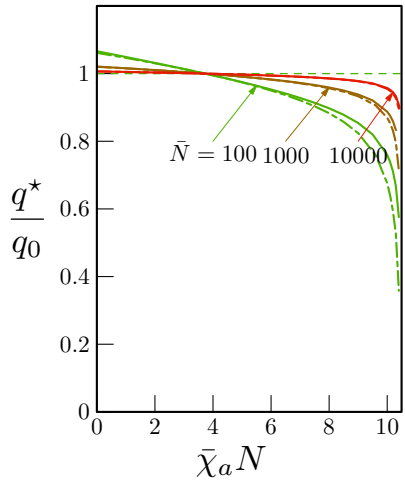}
\caption[]{$\chi N$ dependence of the peak position $q^\star$, normalized by
the RPA theory value $q_0$, for $\bar{N}=100$, 1000 and 10000. Solid lines:
exact results. Dashed lines: estimates using eq. (\ref{eq:fit2}).}
\label{fig:Nfit}
\end{figure}

Let $x_{0}$ be the value of $x$ at which the $f_{0}(x,y)$ is
minimized, so that $x_0 = q_0 R_{g,0} = 1.946$ for a symmetric 
diblock copolymer. Because the RPA prediction $f_{0}(x,y)$
is linear in $y = \chi N$, the minimum with respect to $x$ 
is independent of $y$.  

When the minimum of $f(x,y,\bar{N})$ with respect to $x=qR$ 
remains near $x_{0}$, we may approximate it by Taylor expanding 
both $f_{0}(x,y)$ and $f_{1}(x,y)$ about $x_{0}$ to quadratic 
order in $x - x_0$. By minimizing the resulting approximation
for $f$ with respect to $x$, we obtain
\begin{equation}
  x^\star = x_0 - \frac{f_1'(x_0,y)}{f_0''(x_0,y) + \bar{N}^{-1/2} f_1''(x_0,y)}
  \bar{N}^{-1/2} .
  \label{eq:fit2}
\end{equation}
For symmetric diblock copolymers, $f_{0}''(x_0) = 14.5715$.
Our numerical results for $f_1'(x_0; \chi N)$ and 
$f_1''(x_0; \chi N)$ for symmetric diblock copolymers are
accurately fit by the following empirical formulas:
\begin{align}
  f_1'(x_0, \chi N) &=
  \frac {-54.335 + 14.713 \chi N} {\sqrt{\chi_s N - \chi N}}
  \label{eq:fittingQstar} \\
  f_1''(x_0, \chi N) & =
  \frac {-117.762 + 13.459 \chi N} {\sqrt{\chi_s N - \chi N}}
  \nonumber \\
  & +  \frac {477.324 - 45.569 \chi N} {\chi_s N - \chi N}.
  \label{eq:fittingQstar2nd}
\end{align}
To obtain the self-consistent ROL theory, the parameter $\chi$ 
should be replaced by the self-consistently determined value of
$\barchia N$ throughout the above. The resulting approximation 
for $q^\star$ is compared to results of the full theory in
Fig. \ref{fig:Nfit}. The quality of the fit evidently improve 
as $\bar{N}$ increases, and is adequate for most purposes for 
$\bar{N} \gtrsim 500$. 

% ---------------------------------------------------------------------

% Bibliography
\bibliographystyle{achemso}
\bibliography{}
%\bibliography{reference,dbcFluctTh,dbcFluctEx,bldFluctTh,blendSimulation,diblockSimulation}

\end{document}